\documentclass{article}

\usepackage{arxiv}

\usepackage[utf8]{inputenc} 
\usepackage[T1]{fontenc}    
\usepackage{hyperref}       
\usepackage{url}            
\usepackage{booktabs}       
\usepackage{amsfonts}       
\usepackage{nicefrac}       
\usepackage{microtype}      
\usepackage{graphicx}
\usepackage[square, numbers, comma, sort&compress]{natbib}
\usepackage{doi}
\usepackage{amsmath}
\usepackage{caption}
\usepackage{subcaption}

\title{A Study on the Scattering of Matter Waves through Slits}

\author{
    Hardeep Singh \\
	Department of Physical sciences\\
	UM-DAE Centre for Excellence in Basic Sciences\\
	Mumbai, India \\
	\texttt{hardeep.chhabra18@gmail.com} \\
	\And
    A. Bhagwat \\
	Department of Physical sciences\\
	UM-DAE Centre for Excellence in Basic Sciences\\
	Mumbai, India \\
	\texttt{ameeya@cbs.ac.in} \\
}

\date{}

\renewcommand{\shorttitle}{A Study on the Scattering of Matter Waves through Slits}

\hypersetup{
pdftitle={A Study on the Scattering of Matter Waves through Slits},
pdfsubject={quant-ph},
pdfauthor={Hardeep Singh, A. Bhagwat},
pdfkeywords={Double-slit Experiment, Continuity Equation, Feynman Path Integral, Scattering},
}

\begin{document}
\maketitle

\begin{abstract}
	Scattering of matter waves through slits has been explored using the Feynman Path Integral formalism. We explicitly plot the near-zero probability densities to analyse the behaviour near the slit. Upon doing so, intriguing patterns emerge, most notably the braid-like structure in the case of double slits, whose complexity increases as one increases the number of slits. Furthermore, the plot shows the existence of a transition region, where the distribution of near-zero probability points changes from the braided to the fringe-like structure, which has been analysed by explicitly expressing the wavefunction as a hypergeometric function. These patterns are analysed while considering the continuity equation and its consequences for the regions with zero probability density.
\end{abstract}

\keywords{Double-slit Experiment, Continuity Equation, Feynman Path Integral, Scattering}

\section{Introduction}\label{sec: Introduction}
The famed double-slit experiment, introduced originally by Thomas Young in 
his lectures at the Royal Society in 1802 \cite{young1802}, is unarguably one
of the most beautiful experiments ever performed in the history of 
science. Originally designed to test the corpuscular behaviour of the light 
advocated by Newton, the double slit experiment has been performed in 
the recent times using atoms and electrons to test the Quantum Mechanical
principles \cite{nairz2003quantum}. Originally thought to be impossible, a
version of the experiment
has been performed even with a single electron demonstrating the wave 
properties associated with particles in the microscopic domain
\cite{frabboni2012young}. The 
experiment has taken an important role in the pursuit of understanding 
the Quantum Mechanics itself, one such example being the well - known 
Delayed-choice Experiment \cite{Delayed_Choice}, and as recently by Aharonov in which he used the idea of double slit-experiment to discuss deterministic perspective of quantum mechanics through Heisenberg picture \cite{aharonov2017finally}. Scattering of matter waves composed of large molecules such as $C_{60}$ from multiple slits has been demonstrated experimentally by Anton Zeilinger and collaborators \cite{zeilinger1999wave, zeilinger2003quantum}, which demonstrated that the molecule as a whole act as one quantum object. The double-slit experiment 
thus has a very rich history, yet it is young as ever.

In this paper, we discuss the similar scenario, but with matter waves, which allows us to employ Feynman Path Integral formalism\cite{feynman2010quantum}. This method has been used number of times by various authors to describe scattering through slits \cite{philippidis1979quantum, yabuki1986feynman, gondran2001fentes, sbitnev2009bohmian}. Here, we present an alternate perspective of the analysis of scattering through slits, where we examine the distribution of points having probability densities so small that they can be considered to be zero. As we shall demonstrate in the subsequent sections, this approach yields surprising insight into the behaviour of such systems in the regions near as well as far away from the slit plane.

Quantum Mechanics in conjugation with the equation of continuity, allows one to think of the evolution of a quantum particle following a well defined trajectory, although the precise determination of these trajectories require another set of guiding equations. This idea has been used by Bohm in his formalism of Quantum Mechanics, which is often known as Bohmian Mechanics\cite{Bohm_I}. There exist other formalisms which use the same core idea but follow a different approach to Quantum Mechanics, one prominent example being Quantum-hydrodynamics\cite{wyatt2005quantum}. In this paper, the trajectory picture that has been adopted (Section \ref{sec: Null behaviour}) is assumed to strictly follow the continuity equation.

The paper is organised as follows. A detailed formalism is developed in section \ref{sec: double slit}, along with a few preliminary numerical results. A possible existence of scaling symmetry in the above scenario has been demonstrated there. The notions of Nulls (points with probability density equal to zero) along with Null maps are developed and analysed in section \ref{sec: Null behaviour}. We then present the detailed analytical approach to the present problem including representation of wave-function as hypergeometric functions as well as vectors on the complex plane defined by Fresnel integral is presented in section \ref{sec: analysis}. The summary of this investigation and conclusions are contained in the last section.

\section{Formalism} \label{sec: double slit}

This paper employs Feynman Path Integral approach to study scattering through the slits. There are multiple methods to establish such a scenario. Here, we demonstrate one of such methods by establishing the set up in 1-dimensional case (1-D), as has also been shown in the book by Feynman\cite{feynman2010quantum}.
 
\subsection{One Dimensional Analysis}\label{sec: Scatter 1D}
 Consider a localized particle with mass $m$ at a position $x_0$ at time $t = t_0$. The spatial wave-function of such a particle is given by (see, for example, ref \cite{shankar2012principles}):
 
\begin{equation}\label{eqn: 1D initial}
    \psi\left(x,t=t_0\right) = \delta \left(x-x_0\right)
 \end{equation}
 
The evolution of the wave function of the free particle as per Feynman Path Integral approach with a given initial wave-function $\psi(x,t_0)$ is given as:
\begin{equation}\label{eqn: 1D psi evolution}
    \psi(x_1,t_1) = \int_{-\infty}^{\infty} K(x_1,t_1;x,t_0) \psi(x,t_0) \, dx
 \end{equation}
Where, $K(x_1,t_1;x,t_0)$ is the kernel for the free particle, expressed as (see \cite{feynman2010quantum, Feynman-Path-Paper}):
\begin{equation}\label{eqn: 1D kernel free particle}
    K(x_1,t_1;x,t_0) = \left(\frac{m}{2 \pi \iota \hbar (t_1 - t_0)}\right)^{1/2} \exp \left\{ \frac{\iota m (x_1 - x)^2}{2 \hbar (t_1 - t_0)} \right\} 
\end{equation}
Which upon substitution in Eqn. (\ref{eqn: 1D initial}) yields:
\begin{equation}
    \psi(x_1,t_1) =\left(\frac{m}{2 \pi \iota \hbar (t_1 - t_0)}\right)^{1/2} \exp \left\{ \frac{\iota m (x_1 - x_0)^2}{2 \hbar (t_1 - t_0)} \right\}
\end{equation} 
In the case of 1-D, slits can be introduced at a given time $t_1$, and only the points present inside the slit are expected to contribute to the wave-function evolution beyond. If slit points are denoted by set $S_0$ then the wave-function at a later space-time point becomes:
\begin{equation}
    \psi(x_2,t_2) = \int_{S_0} K(x_2,t_2;x_1,t_1) \psi(x_1,t_1) \,dx_1
\end{equation}
Using Eqn. (\ref{eqn: 1D kernel free particle}), the wave-function takes the following form:
\begin{equation} \label{eq:psi}
    \psi(x_2, t_2) = \int_{S_0}\left(\frac{m}{2 \pi \iota \hbar t^{\prime}}\right)^{1/2} \exp\left\{\frac{\iota m D^{\prime}}{2\hbar t^{\prime}}\right\}
    \left(\frac{m}{2 \pi \iota \hbar t^{\prime\prime}}\right)^{1/2} \exp\left\{\frac{\iota m D^{\prime\prime}}{2\hbar t^{\prime\prime}}\right\} \,dx_1
\end{equation}

with $D^{\prime} = (x_1 - x_0)^2$, $D^{\prime\prime} = (x_2 - x_1)^2$, $t^{\prime} = t_1 - t_0$ and $t^{\prime\prime} = t_2 - t_1$. Since, quantity $2 \pi \hbar t/m$ has the dimensions $L^2$, it allows one to express this quantity as a product of two entities one being a constant and the other being a variable, each with the dimension of length. With this motivation, we express the time variables $t^{\prime}$ and $t^{\prime\prime}$ in terms of the constant $\lambda$ and variables $z^{\prime}$ and $z^{\prime\prime}$ (both having dimensions of length) as:

\begin{equation}\label{eqn: 1d_substitution }
    t^{\prime} = \frac{\lambda m z^{\prime}}{2 \pi \hbar}\,\,;\,\,
    t^{\prime\prime} = \frac{\lambda m z^{\prime\prime}}{2 \pi \hbar}
\end{equation}
Which simplifies the equation (\ref{eq:psi}) to:
\begin{equation} \label{eqn: 1D final}
    \psi(x_2, t_2) = \frac{1}{\iota \lambda} \int_{S_0} \frac{1}{\sqrt{z^{\prime}\,z^{\prime\prime}}} \exp \left\{ \frac{\iota \pi}{\lambda} \left(\frac{D^{\prime}}{z^{\prime}} + \frac{D^{\prime\prime}}{z^{\prime\prime}}\right)\right\}\,dx_1
\end{equation}
The equation eqn (\ref{eqn: 1D final}) can easily be used for numerical analysis.

\subsection{Two Dimensional Analysis}
By similar set of arguments as discussed in previous subsection, one can attempt to write the wave-function in 2-dimensional case (2-D).\\
For the initial state, consider a collapsed state:
\begin{equation}
    \psi\left(x,y,t=t_0\right)=\delta\left(x-x_0\right)\delta\left(y-y_0\right)
\end{equation}
The kernel in 2-D is slightly different from that in 1D. By taking Lagrangian of free particle in 2-D i.e. 
\begin{equation}
    \mathcal{L} = \frac{1}{2} \, m(\dot{x}^2 + \dot{y}^2)
\end{equation}
and following the procedure as given in \cite{feynman2010quantum, shankar2012principles}, one obtains the following Kernel for a particle going from position $(x_a, y_a)$ at time $t_a$ to $(x_b, y_b)$ at time $t_b$ :
\begin{equation}\label{eqn: kernel 2D }
    K\left(b,a\right)=\frac{m}{2\pi\iota\hbar\left(t_b-t_a\right)}
    \exp{\frac{\iota m\left[\left(x_b-x_a\right)^2+\left(y_b-y_a\right)^2\right]}{2\hbar\left(t_b-t_a\right)}}
\end{equation}
Introducing a slit at a specific value of $y$, let us choose it to be $y=y_1$, then at the position of the slits, the wave-function will be
\begin{equation}\label{eqn: wavefunction 2D}
    \psi\left(x_1,y_1,t_1\right)=\frac{m}{2\pi\iota\hbar\left(t_1-t_0\right)}
    \exp{\frac{\iota m\left[\left(x_1-x_0\right)^2+\left(y_1-y_0\right)^2\right]}{2\hbar\left(t_1-t_0\right)}}
\end{equation}
It is not straightforward to compute the evolution beyond the slit in higher dimensional cases. One reason is that one has to consider the paths contributed from the wave-function present before the slit, and there is a possibility of backflow of paths as well as looped paths around the slits \cite{yabuki1986feynman}. In order to understand the salient feature of the scenario, the analysis can be carried out by introducing certain approximations, leading to a considerable simplification of the analysis. 

Note that the initial condition (collapsed state) is represented by Dirac delta function, by the uncertainty principle the wave-function is in the superposition of all the possible momentum states. However, our desired state of analysis in 2-D is a plane-wave having a specific momentum $p$ corresponding to wavelength $\lambda$, given by the de Broglie relation $p\ =\ h/\lambda$.
One proposed solution is to slice the wave-function originating from the collapsed state only to select the part which has the effective wavelength $\lambda$. Consider the kernel of the free particle:
\begin{equation}
    K=F\left(t\right)\exp{\frac{\iota m y^2}{2\hbar t}}
\end{equation}
$F(t)$ is the normalization function, which depends on the dimensions of the space in the system. To find the effective wavelength, we propose to increment $y$ by $\lambda$. This increment should introduce a phase difference of $2\pi$ in the argument of the complex exponential. Mathematically it is expressed as:
\begin{equation}
    2\pi=\frac{m\left(y+\lambda\right)^2}{2\hbar t}-\frac{my^2}{2\hbar t}=\frac{my\lambda}{\hbar t}+\frac{m\lambda^2}{2\hbar t}
\end{equation}
If one goes sufficiently far away from the point of origin such that $y\ \gg\ \lambda$, then one can essentially ignore $\lambda^2$ contribution. Therefore:
\begin{equation}\label{eqn: lambda 2D }
    \lambda=\frac{2\pi\hbar}{m\left(y/t\right)}
\end{equation}
We can interpret this form as following; If one takes the ratio $y/t$ as constant, one is essentially tracing the sliced wave-function whose effective wavelength is $\lambda$. Therefore, we look only at those paths that conserve this ratio. This idealization allows us to deal with the issue of backflow, as the selected paths encounter the slit only once at specific $y/t$. Using this ratio, we can make the substitution.
\begin{equation}\label{eqn: 2d_substitution }
    t=\frac{\lambda m y}{2\pi\hbar}
\end{equation}
The above relation strikingly resembles the dimensional substitution we made in 1-D (Eqn. (\ref{eqn: 1d_substitution })).
Thus, we can proceed to write the evolution wave-function beyond the slit at the point $\left(x_2,y_2,t_2\right)$.
\begin{equation}\label{eqn: propagator wave 2D}
    \psi\left(x_2,y_2,t_2\right)=\int_{S^\prime} K\left(b,a\right)\psi\left(x_1,y_1,t_1\right)\, dx_1
\end{equation}

The kernel $K(b, a)$ and wave function $\psi(x_1,y_1,t_1)$ appearing here in this expression has been explicitly define in Eqns. (\ref{eqn: kernel 2D }) and (\ref{eqn: wavefunction 2D}) respectively, and $S^{\prime}$ is the set of points lying inside the slit.
Upon changing the variables as $x^\prime=x_1-x_0$, $y^\prime=y_1-y_0$, $t^\prime=t_1-t_0$, $x^{\prime\prime}=x_2-x_1$, $y^{\prime\prime}=y_2-y_1$ and $t^{\prime\prime}=t_2-t_1$, Eqn. (\ref{eqn: propagator wave 2D}) becomes; 
\begin{equation}
    \psi\left(x_2,y_2,t_2\right)=\int_{S^\prime}F\left(t^\prime\right)\exp{\frac{\iota m\left[x^{\prime2}+y^{\prime2}\right]}{2\hbar t^\prime}}
    F\left(t^{\prime\prime}\right)\exp{\frac{\iota m\left[x^{\prime\prime2}+y^{\prime\prime2}\right]}{2\hbar t^{\prime\prime}}}\, dx_1
\end{equation}
Now, making use of the fact that $y/t$ is a constant (Eqn. \ref{eqn: lambda 2D }), we get:
\begin{equation}
    \psi\left(x_2,y_2,t_2\right)=\Tilde{F}\left(y^\prime\right)\Tilde{F}\left(y^{\prime\prime}\right)\exp{\frac{\iota\pi\left[y^\prime+y^{\prime\prime}\right]}{\lambda}}
    \int_{S^\prime} \exp{\frac{\iota\pi}{\lambda}\left(\frac{x^{\prime2}}{y^\prime}+\frac{x^{\prime\prime2}}{y^{\prime\prime}}\right)}\, dx_1
\end{equation}

The factors $\Tilde{F}\left(y^\prime\right)$ and $\Tilde{F}\left(y^{\prime\prime}\right)$ are obtained by transforming $F\left(t^\prime\right)$ and $F\left(t^{\prime\prime}\right)$ respectively, under the transformation $t\ \mapsto y$. Further substituting
\begin{equation*}
    T\left(y^\prime,y^{\prime\prime}\right)=\Tilde{F}\left(y^\prime\right)\Tilde{F}\left(y^{\prime\prime}\right)\exp{\frac{\iota\pi\left[y^\prime+y^{\prime\prime}\right]}{\lambda}}\, ,   
\end{equation*}
the wave-function gets the similar form as was in the case of 1-D. 
\begin{equation} \label{eqn: idealized 2D }
    \psi\left(x_2,y_2,t_2\right)=T\left(y^\prime,y^{\prime\prime}\right)\int_{S^\prime} e x p{\left\{\frac{\iota\pi}{\lambda}\left(\frac{x^{\prime2}}{y^\prime}+\frac{x^{\prime\prime2}}{y^{\prime\prime}}\right)\right\}}\, dx_1
\end{equation}
This suggests that the shape of probability density obtained in the idealized 2-D case is identical to that in 1-D.

\subsection{Steady-State Analysis of the 2-D Case} 
Although there is no time variable on the right-hand side of the Eqn. (\ref{eqn: idealized 2D }), it is not a steady-state solution. 
In order to introduce a steady-state, one needs to have a continuously emitting source. Which can be implemented through a boundary condition as follows
\begin{equation}
    \psi_0\left(x,y_0,t\right)=\delta\left(x-x_0\right)\exp{\frac{\iota E t}{\hbar}}
\end{equation}
Here, $E$ is the energy of the non interacting particles that are introduced into the system.

With this modification, the source is continuously injecting particles in the system. Therefore, to compute the probability density at a point in space, it is required to consider the $\psi$ introduced at previous times as well. To simplify the analysis we employ the time slicing method, where the entire time-interval $[0,\infty)$ has been sliced into countable number of bins, each with a fixed size $\epsilon$.

Let $l\epsilon$ be the time elapsed since a wave-function has been introduced to the system, where $l$ is a positive integer. For the sake of brevity, we use $D=x^{\prime2}+y^{\prime2}$, here the symbols have the usual meaning as defined before. Hence we can write the steady-state wave-function as: 

\begin{equation}\label{eqn: steady state sum }
    \psi\left(D\right)=\sum_{l=1}^{\infty}\psi\left(D,l\epsilon\right)
\end{equation}
where for a free particle,
\begin{equation}
    \psi\left(D,l\epsilon\right)=F\left(l\epsilon\right)\exp{\frac{\iota m D}{2\hbar l\epsilon}}\exp{\frac{\iota E l\epsilon}{\hbar}}
\end{equation}
Noting that $F\left(l\epsilon\right)\propto 1 / l \epsilon$, one can effectively ignore wave-functions for which time elapsed is large such that $F(l\epsilon) \approx 0$.

Consider a wave-function for which time elapsed is $t^\prime$ and look at the wave-function from neighboring times, i.e. $t^\prime\pm\Delta t$, where $\Delta t\ll t^\prime$. These are given by:
\begin{equation}\label{eqn: psi d' t'}
    \psi\left(D,t^\prime\right)=F\left(t^\prime\right)\exp{\frac{\iota m D}{2\hbar\left(t^\prime\right)}}\exp{\frac{\iota E\left(t^\prime\right)}{\hbar}}\\
\end{equation}    
\begin{equation}
    \psi\left(D,t^\prime\pm\Delta t\right)=F\left(t^\prime\pm\Delta t\right)\exp{\frac{\iota m D}{2\hbar\left(t^\prime\pm\Delta t\right)\ }}
    \exp{\frac{\iota E\left(t^\prime\pm\Delta t\right)}{\hbar}}
\end{equation}
Since $\Delta t\ll t^\prime$, we can write
\begin{equation*}
    \frac{1}{t^\prime\pm\Delta t}\simeq\frac{1\mp\left(\Delta t/t^\prime\right)}{t^\prime}\, .
\end{equation*}
Given this and the fact that $F$ is inversely proportional to $t$ we can write 
\begin{equation*}
    F\left(t^\prime\pm\Delta t\right)\simeq F\left(t^\prime\right)\left[1\mp \left(\Delta t/t^\prime\right)\right] \, .    
\end{equation*}
Using these approximations, one obtains:
\begin{equation}
    \psi\left(D,t^\prime\pm\Delta t\right)=F\left(t^\prime\right)\left[1\mp \left(\Delta t/t^\prime\right)\right]
    \exp \left\{\frac{\iota m D}{2 \hbar t^\prime}\left[1\mp (\Delta t/ t^\prime)\right]\right\}\exp\left\{\frac{\iota E t^\prime}{\hbar}\left[1\pm(\Delta t/ t^\prime)\right]\right\}
\end{equation}
After a simple rearrangement of terms,we get:
\begin{equation}
    \psi\left(D,t^\prime\pm\Delta t\right)=F\left(t^\prime\right)\exp{\frac{\iota m D}{2\hbar t^\prime}}\exp{\frac{\iota E t^\prime}{\hbar}}
    \left[1\mp \left(\Delta t/t^\prime\right)\right] \exp{\frac{\mp\iota\left(\Delta t/t^\prime\right)}{\hbar}\left[\frac{m D}{2t^\prime}-Et^\prime\right]}
\end{equation}
Evidently it can be re-expressed using Eqn. (\ref{eqn: psi d' t'}) as:
\begin{equation}\label{eqn: steady 2D }
    \psi\left(D,t^\prime\pm\Delta t\right)=\psi\left(D,t^\prime\right)\ \left[1\mp \left(\Delta t/t^\prime\right)\right]
    \exp{\frac{\mp\iota\left(\Delta t/t^\prime\right)}{\hbar}\left[\frac{mD}{2t^\prime}-Et^\prime\right]}
\end{equation}
Recall that $E$ is the energy of the particles emitted by the source and is a constant. In the classical scenario, it is given as $mv^2/2$. Furthermore, $v$ can be taken as the constant ratio of position and time variable. If we take $v$ to be $y^\prime/t^\prime$, then it is evident to write down $Et^\prime=my^{\prime2}/2t^\prime$. Under these conditions, Eqn. (\ref{eqn: steady 2D }) can be written as:
\begin{equation}
    \psi\left(D,t^\prime\pm\Delta t\right)=\psi\left(D,t^\prime\right)\left[1\mp \left(\Delta t/t^\prime\right)\right]
    \exp{\frac{\mp\iota\left(\Delta t/t^\prime\right)}{\hbar}\frac{mx^{\prime2}}{2t^\prime}}
\end{equation}
If the argument inside the exponent is taken to be small as well, given $x^\prime$ is comparable with $y^\prime$, then it is possible to write:
\begin{equation}
    \psi\left(D,t^\prime\pm\Delta t\right)=\psi\left(D,t^\prime\right)\left[1\mp \left(\Delta t/t^\prime\right)\right]
    \left[1\mp\frac{\iota\left(\Delta t/t^\prime\right)}{\hbar}\frac{mx^{\prime2}}{2t^\prime}\right]
\end{equation}
Expanding the square brackets and retaining the terms only up to first-order in $\left(\Delta t/t^\prime\right)$, one gets:
\begin{equation}
    \psi\left(D,t^\prime\pm\Delta t\right)=\psi\left(D,t^\prime\right)\left[1\mp\left(1+\frac{\iota m x^{\prime2}}{2\hbar t^\prime}\right)\frac{\Delta t}{t^\prime}\right]
\end{equation}
Therefore,
\begin{equation}
    \psi\left(D,t^\prime+\Delta t\right)+\psi\left(D,t^\prime-\Delta t\right)\approx2\psi\left(D,t^\prime\right)
\end{equation}
Hence, we can write using Eqn. (\ref{eqn: steady state sum }):
\begin{equation}
    \psi\left(D\right)\approx k\psi\left(D,t^\prime\right)\ 
\end{equation}

With the condition that $y^\prime/t^\prime=$ const, such that they are related to energy as, $E=my^{\prime2}/2t^{\prime2}$.
Therefore, one concludes that the idealization of the 2-D case that has been developed in the previous sub-section is a good approximation to the steady state scenario. Henceforth, we shall restrict our analysis to the 1-D case.

\begin{figure}
    \centering
    \includegraphics[width = 0.8\textwidth]{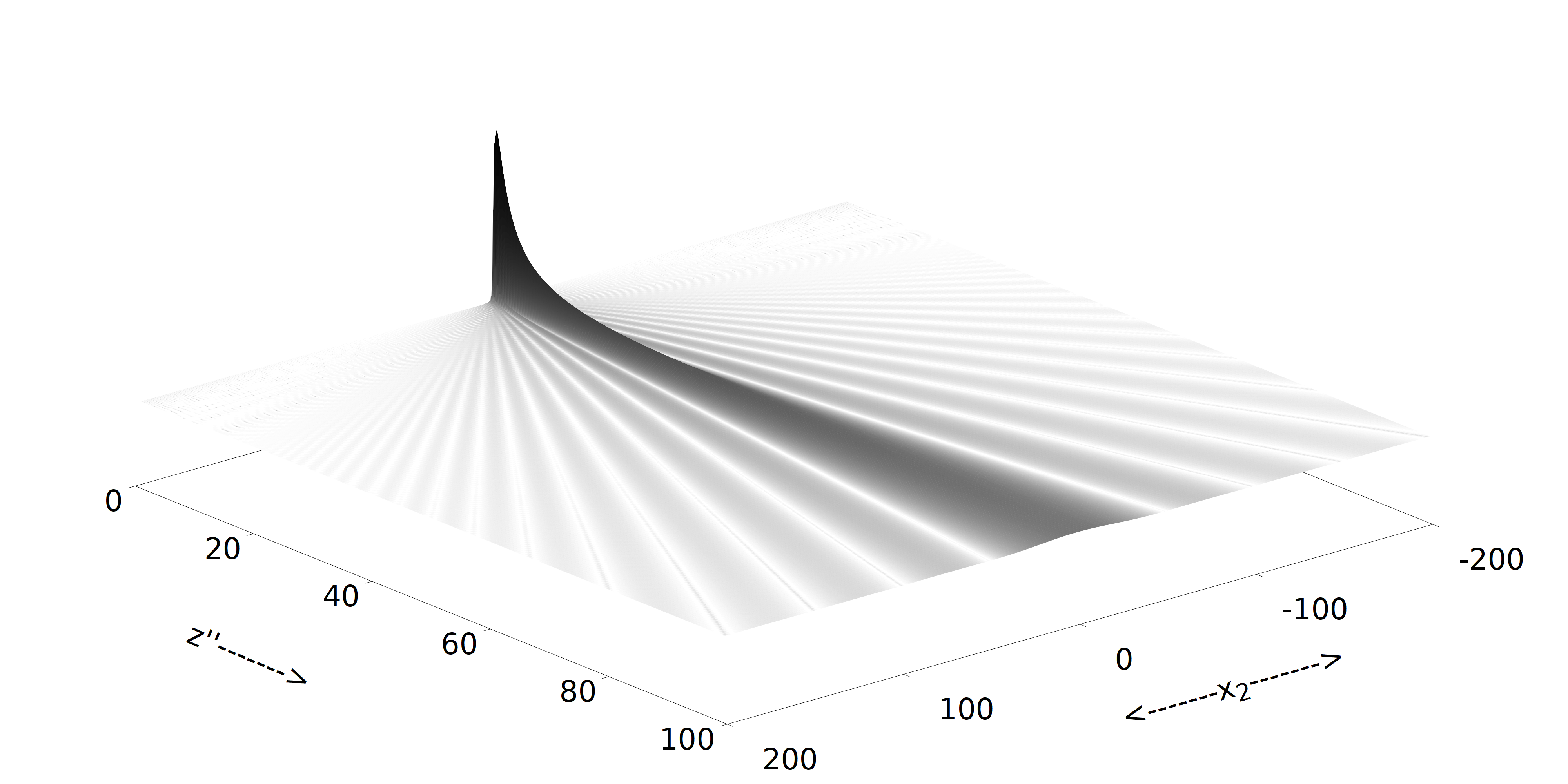}
    \caption{Evolution of the probability density through a single slit present at x = 0 with slit width 2$\lambda$.}
    \label{fig: single slit 2 }
\end{figure}

\begin{figure}
    \centering
    \includegraphics[width = 0.8\textwidth]{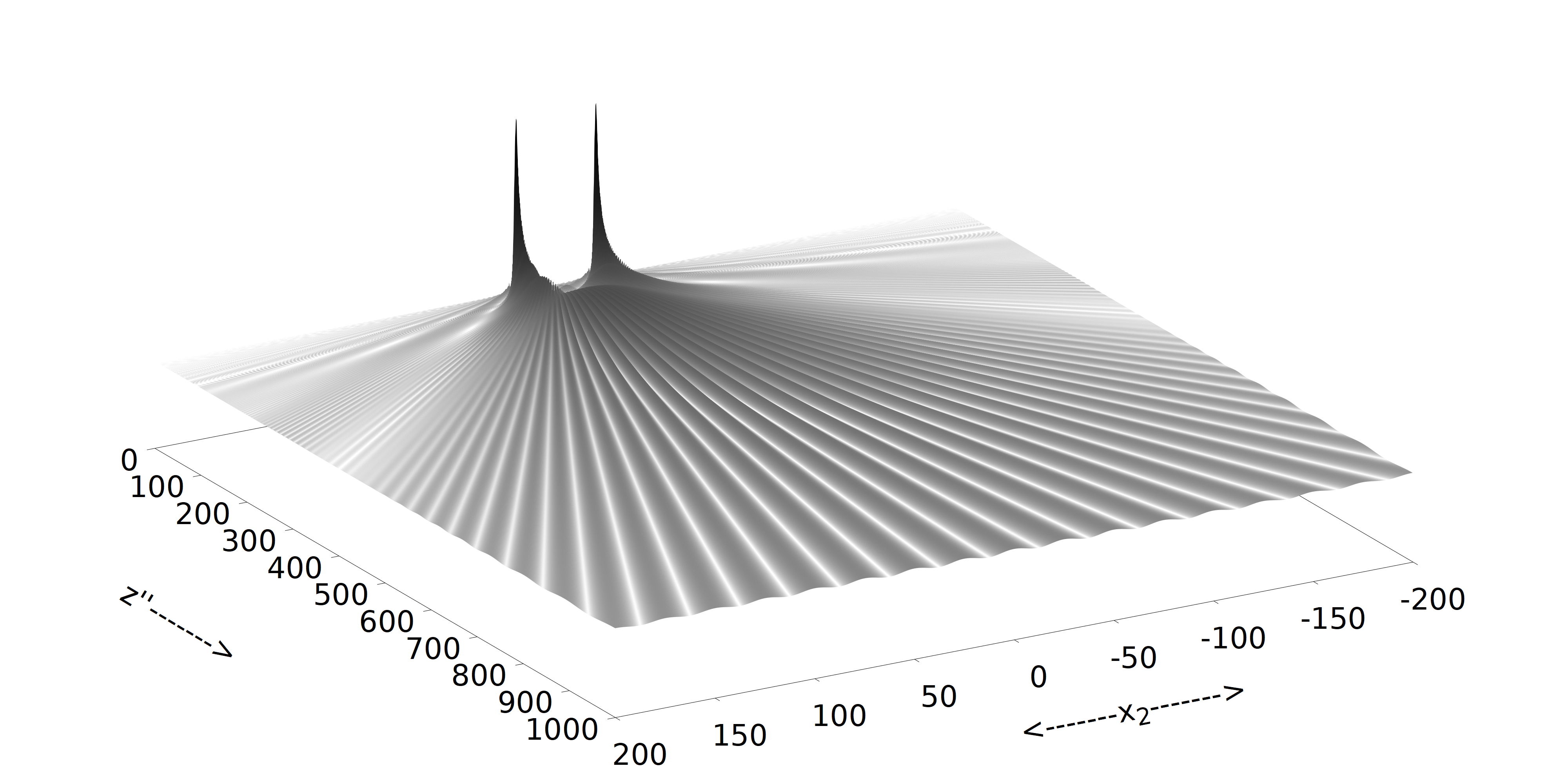}
    \caption{Evolution of the probability density through the double slit present at x = -20 and x = 20}
    \label{fig:2slits}
\end{figure}

\subsection{Preliminary Numerical Results}
As all the parameters in Eqn. (\ref{eqn: 1D final}) have either the dimensions of $L$ or $L^2$, we express all the parameters in the units of $\lambda$ and $\lambda^2$ respectively. For the sake of convenience, $\lambda$ is chosen to be 1. The factor $1/\iota$ is omitted from the expression as well, as a constant phase does not impact the probability density at any given point. The simplified expression thus obtained reads:

\begin{equation}\label{eqn: integral pre simplification }
    \psi\left(x_2,t_2\right)=\frac{1}{\sqrt{z^\prime z^{\prime\prime}}}\int_{S^\prime} e x p{\left\{\iota\pi\left(\frac{x^{\prime2}}{z^\prime}+\frac{x^{\prime\prime2}}{z^{\prime\prime}}\right)\right\}}\, dx_1
\end{equation}

We first demonstrate the diffraction only through a single slit. One expects to have characteristic central global maximum followed by local maxima on either sides. The calculated probability densities plotted in Fig. \ref{fig: single slit 2 } agrees with the well known evolution of the single slit pattern.

Similarly, in the case of double slits, the plot of double slit interference has been demonstrated in Fig. \ref{fig:2slits}, which depicts the characteristic evolution of the double slit pattern.

Inspection of these two figures reveals that the two patterns which have been shown above are as expected, which gives a fair confidence in the validity of the numerical procedures adopted here.

\begin{figure*}
    \centering
    \begin{subfigure}{0.49\textwidth}
        \centering
        \includegraphics[width = \textwidth]{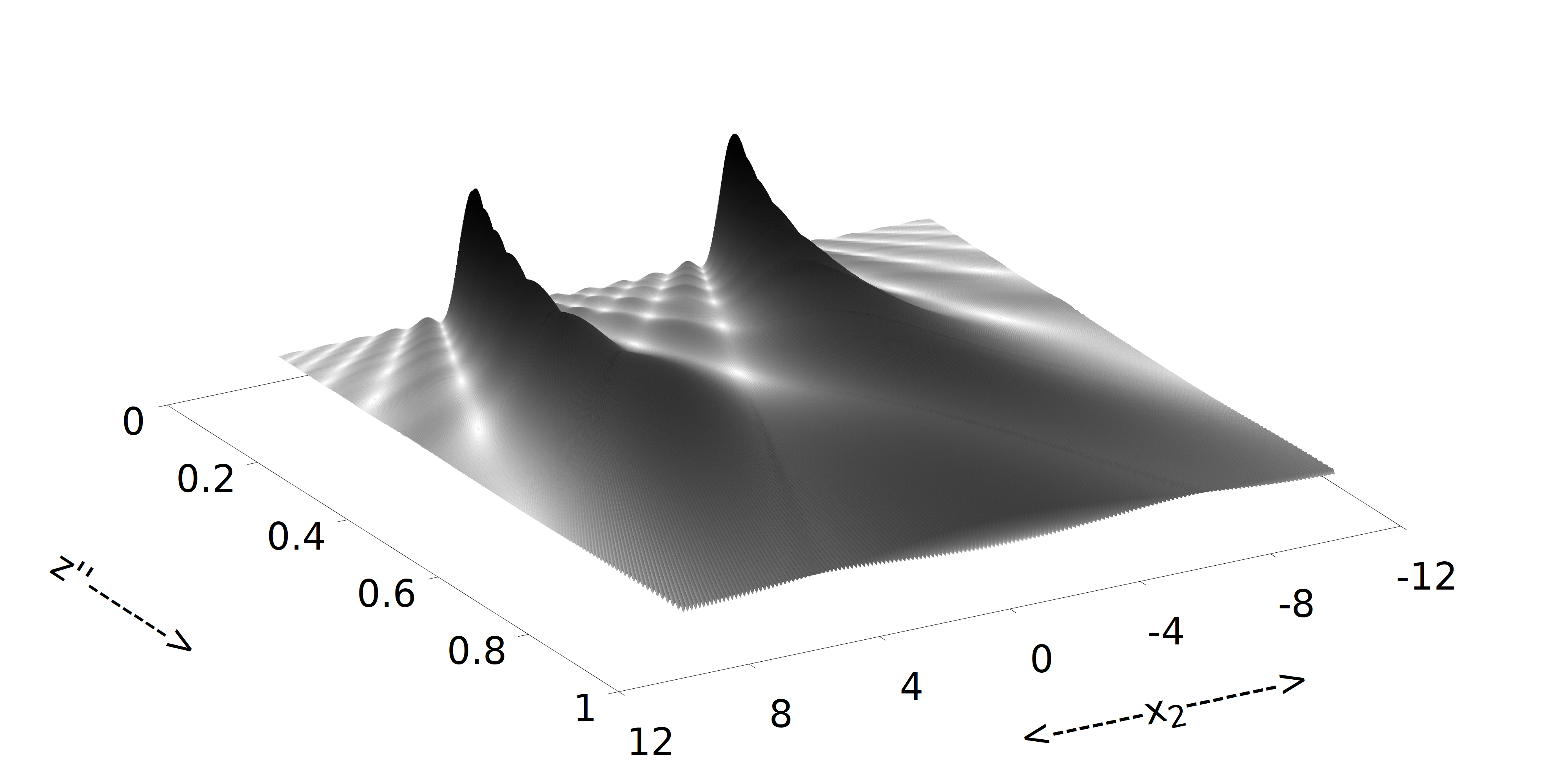}
        \caption{Slit Width $= 0.1\lambda$; Inter Slit Distance $= 8 \lambda$ }
    \end{subfigure}
    \hfill
    \begin{subfigure}{0.49\textwidth}
        \centering
        \includegraphics[width = \textwidth]{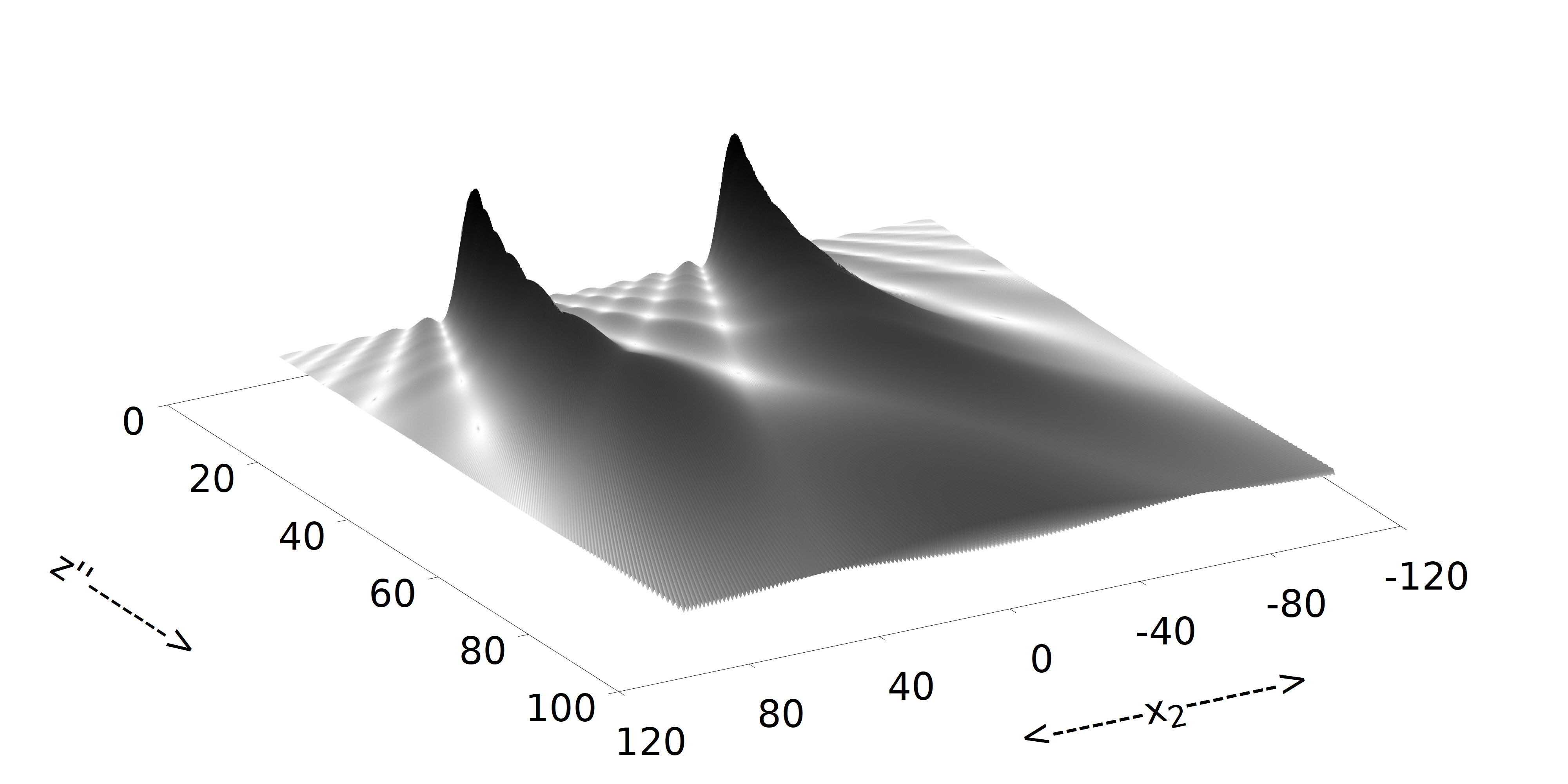}
        \caption{Slit Width $= 1\lambda$; Inter Slit Distance $= 80 \lambda$}
    \end{subfigure}
    
    \begin{subfigure}{0.49\textwidth}
        \centering
        \includegraphics[width = \textwidth]{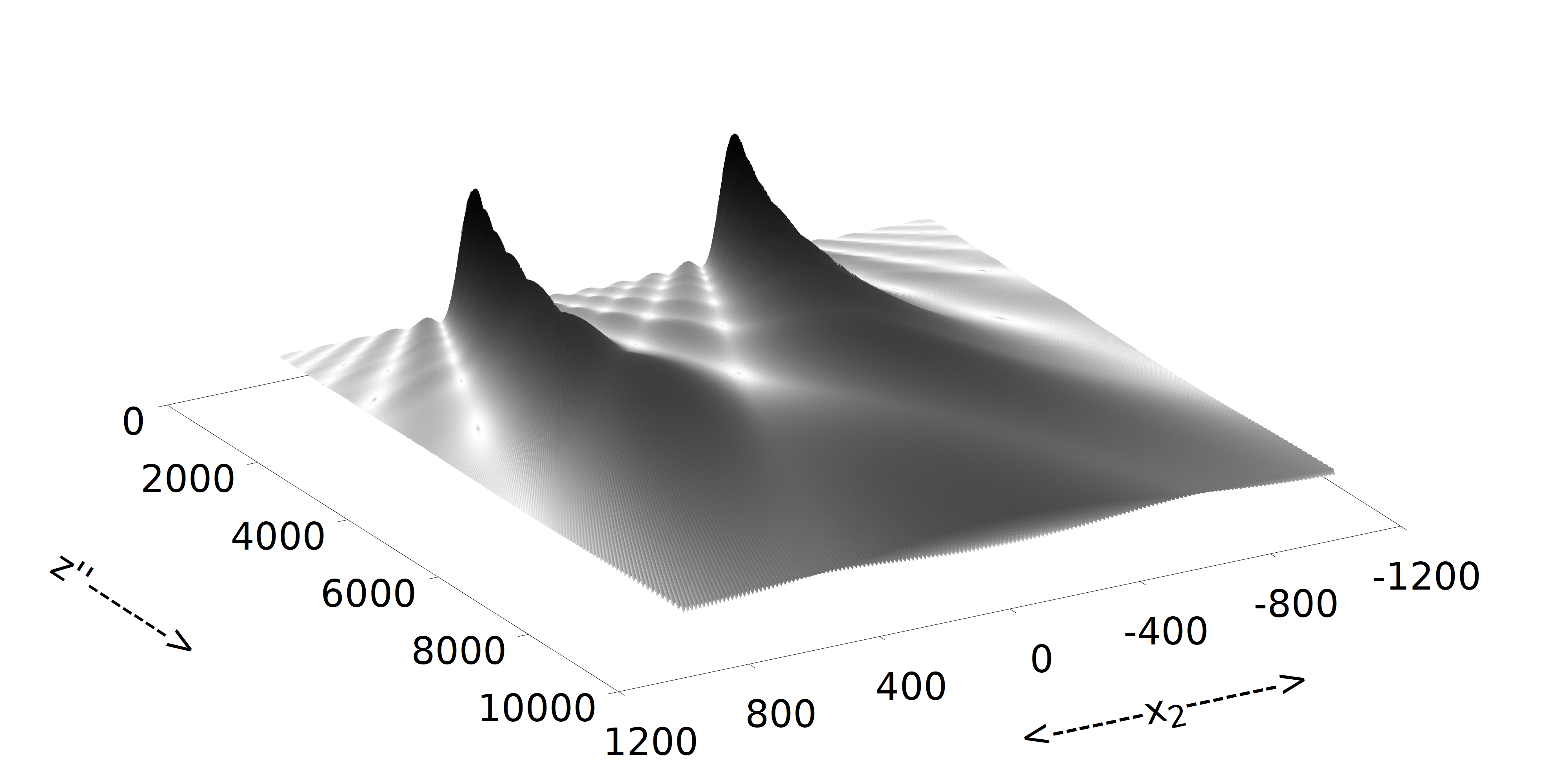}
        \caption{Slit Width $= 10\lambda$; Inter Slit Distance $= 800 \lambda$}
    \end{subfigure}
    \hfill
    \begin{subfigure}{0.49\textwidth}
        \centering
        \includegraphics[width = \textwidth]{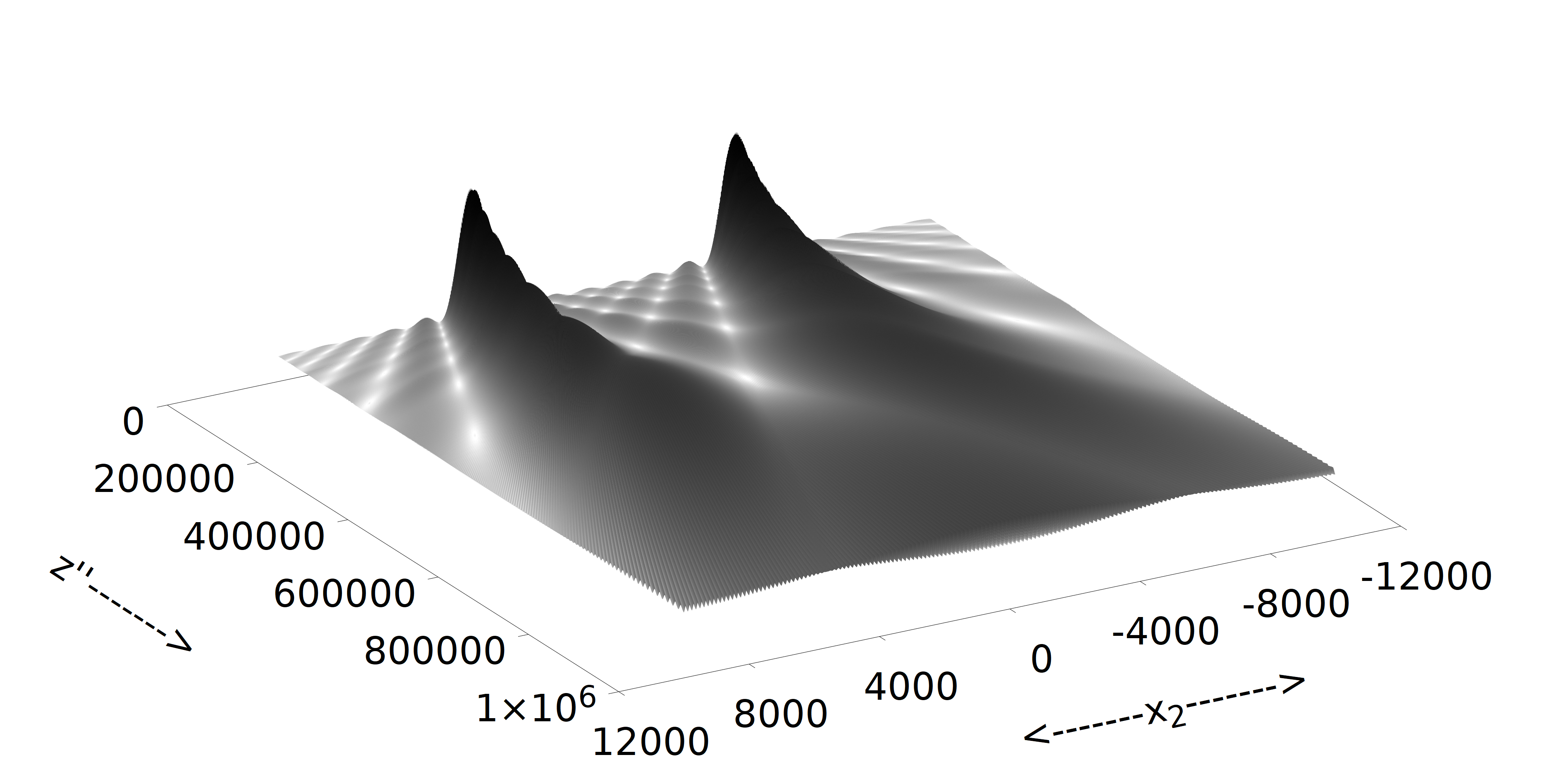}
        \caption{Slit Width $= 100\lambda$; Inter Slit Distance $= 8000 \lambda$}
    \end{subfigure}
    
    \caption{In the set of figures above ratio between Slit Width and the Inter Slit Distance is preserved. It is apparent from figures A to D that $z^{\prime\prime}\,:\,{x_2}^2\,:\, SW^2$ is preserved as the consequence.}
    \label{fig: scaling symmetry}
\end{figure*}

\subsection{Scaling Symmetry}\label{sec: scaling symmetry}

Here we make an approximation that $ z^\prime\gg x^\prime$. Since, $z^\prime$ is a function of $t^\prime$ (see Eqn. \ref{eqn: 1d_substitution }), the above condition is equivalent to  $t^\prime\rightarrow\infty$. Thus the first term inside the exponent in Eqn. \ref{eqn: integral pre simplification } can be dropped. Further, through the evolution beyond the slit plane, $ z^\prime $ is a constant. Thus, it can be safely omitted from the factor outside the integral, without affecting the shape of the wave-function. Therefore, the simplified form of the integral becomes:

\begin{equation}\label{eqn: 1d infinity source }
\psi\left(x_2,t_2\right)=\frac{\ 1}{\sqrt{z^{\prime\prime}}}\int_{S^\prime} e x p{\left\{\iota\pi\frac{x^{\prime\prime2}}{z^{\prime\prime}}\right\}}\, dx_1
\end{equation}
Scaling symmetry can directly be observed from this form. On preserving the ratio between slit-width ($SW$) and inter-slit distance ($ISD$), the set $S^\prime$ can be defined only with $SW$ and $ISD$, thus scaling them is equivalent to scaling $x_1$. Therefore, the form of the integrand is preserved if one scales $z^{\prime\prime}$ as $x^{\prime\prime2}$. However, recall that $x^{\prime\prime2}=x_2-x_1$, which indicates that if $x_2$ is scaled as $x_1$, then $x^{\prime\prime2}$ is scaled as square of the chosen scaling factor. Thus, if one wishes to preserve the shape of the probability density, then $z^{\prime\prime}$ must scale as ${x_2}^2$.

To summarize the argument, if ratio $SW\,:\, ISD$ is preserved, then $z^{\prime\prime}\,:\,{x_2}^2\,:\, SW^2$ should also be preserved. This symmetry has been demonstrated numerically in Fig. (\ref{fig: scaling symmetry}).

It is important to note that this scaling symmetry is independent of the $\lambda$. Thus, it can help extensively to study the close slit structures in macro scales. As we can essentially scale the apparatus thus magnifying the structures and making it experimentally perceivable.

\section{Behaviour of Null}\label{sec: Null behaviour}

Zero probability density points which we shall call Null became interesting points to study. It is natural to wonder how the Null points or boundaries might be forming in scattering through slits. It is known that the middle of the screen is the maximum, which has minima on either side. These minima seemingly appear to be Null; if they extend through the slit plane, then it is only natural to ponder can a trajectory be constructed which can connect from the slit to the central maximum. With this inspiration, it felt only appropriate that we should plot and analyze the points in space that are Null. Before proceeding allow us to analyze some properties akin to Null.

\subsection{Properties of Null}\label{property trap traj}
Any trajectory picture where the particle number is conserved must satisfy the continuity equation of the following form:
\begin{equation}
    \frac{\partial\rho}{\partial t}=-\mathbf{\nabla}\cdot\rho\mathbf{v}
\end{equation}
As the model used for the study is in 1-D. Thus, the appropriate continuity equation will be:
\begin{equation}
    \frac{\partial\rho}{\partial t}=-\frac{\partial\left(\rho v\right)}{\partial x}
\end{equation}
Integrating over the domain with the boundary of zero probability density:
\begin{equation}\label{eqn: continuity rho v }
    \int_{x_1}^{x_2}{\frac{\partial\rho}{\partial t}\ }dx=-\int_{x_1}^{x_2}{\frac{d\left(\rho v\right)}{d x}\ }dx=-\rho v\mid_{x_1}^{x_2}\ =\ 0 
\end{equation}
Here $x_2$ and $x_1$ are the boundaries, and in the case of dynamical boundaries, they are the functions of $t$. Using Leibniz integral rule:
\begin{equation}
    \begin{split}
        \frac{d}{d t}\left(\int_{x_1}^{x_2}\rho\left(x,t\right)\, d x\right)
        & =\rho\left(x_2,t\right)\cdot\frac{d x_2}{d t}\\
        & -\rho\left(x_1,t\right)\cdot\frac{d x_1}{\partial t}\\
        & +\int_{x_1}^{x_2}\frac{\partial\rho\left(x,t\right)}{\partial t}\, dx
    \end{split}
\end{equation}
As $x_1$ and $x_2$ is defined such that $\rho$ is zero on them at all times, i.e., $\rho\left(x_1,t\right)$ and $ \rho\left(x_2,t\right) $ are zero. Therefore, even if $ x_1$ and $ x_2 $ are functions of time; the first two terms on the right-hand side will be zero.\\
Hence,
\begin{equation}
    \int_{x_1}^{x_2}\frac{\partial\rho}{\partial t}\, dx=\frac{d}{d t}\left(\int_{x_1}^{x_2}\rho\, d x\right)=0
\end{equation}
This relation implies that the total probability density enclosed by the boundary having zero probability density remains unchanged. In other words, the trajectories are trapped within.\\
In the domain of quantum mechanics, one has the notion of the continuity equation derivable from the Schr\"odinger's equation, which is given as:
\begin{equation}
    \frac{\partial\rho}{\partial t}=-\mathbf{\nabla}\cdot\left(\frac{\hbar}{2m\iota}\left[\psi^\ast \mathbf{\nabla}\left(\psi\right)-\psi\mathbf{\nabla}\left(\psi^\ast\right)\right]\right)
\end{equation}
Where, $\frac{\hbar}{2m\iota}\left[\psi^\ast\mathbf{\nabla}\left(\psi\right)-\psi\mathbf{\nabla}\left(\psi^\ast\right)\right] = \mathbf{j}$ is the probability density current.

To get further insight into the probability density current, we write the wave-function in polar form:
\begin{equation}\label{eqn: polar }
    \psi=Re^{\iota S/\hbar}
\end{equation}
Both $R$ and $S$ are real functions. The quantity $S$ can be written as:
\begin{equation}
    S=-\frac{\iota\hbar}{2}ln{\left(\frac{\psi}{\psi^\ast}\right)}
\end{equation}
Upon taking the gradient of $S$, one obtains:
\begin{equation}
    \mathbf{\nabla} S=-\frac{\iota\hbar}{2}\left[\frac{\mathbf{\nabla}\psi}{\psi}-\frac{\mathbf{\nabla}\psi^\ast}{\psi^\ast}\right]
\end{equation}
Multiplying with $\rho$ on both sides, we get:
\begin{equation}
    \rho\mathbf{\nabla} S=-\frac{\iota\hbar}{2}\left[\psi^\ast\mathbf{\nabla}\psi-\psi\mathbf{\nabla}\psi^\ast\right]\ = m \mathbf{j}
\end{equation}
Therefore in the trajectory picture, $\mathbf{\nabla} S/m$ can be identified with velocity of particles along the possible trajectories and $\mathbf{\nabla} S$ as the momentum of the particle \cite{Bohm_I}.
Given this, scattering through slits present an interesting case to study. We are aware that far away from the slit, the probability density distribution has consecutive maxima and minima. If one assumes the minima correspond to the Null, then one essentially knows that the particle is trapped between the consecutive minima, and there are multiple such traps. However, the number of slits are limited. Thus one can pose the question, how a particle trajectory must have led the particle inside the trap, and how Null points evolve beyond the slits.

\section{Null Maps}
In this section we present the plots where the loci of Nulls are marked. It has been shown in the previous section that they have an interesting consequence for the trajectory picture. We shall call these loci as Null Maps. In the maps presented in this study, the points where the probability density falls below $10^{-14}$ have been marked. To speed up the process, Monte-Carlo sampling has been employed on a CUDA platform\cite{sanders2010cuda}, in which points are randomly generated using the standard C library function.

\begin{figure}
    \centering
    \includegraphics[width = 0.8\textwidth]{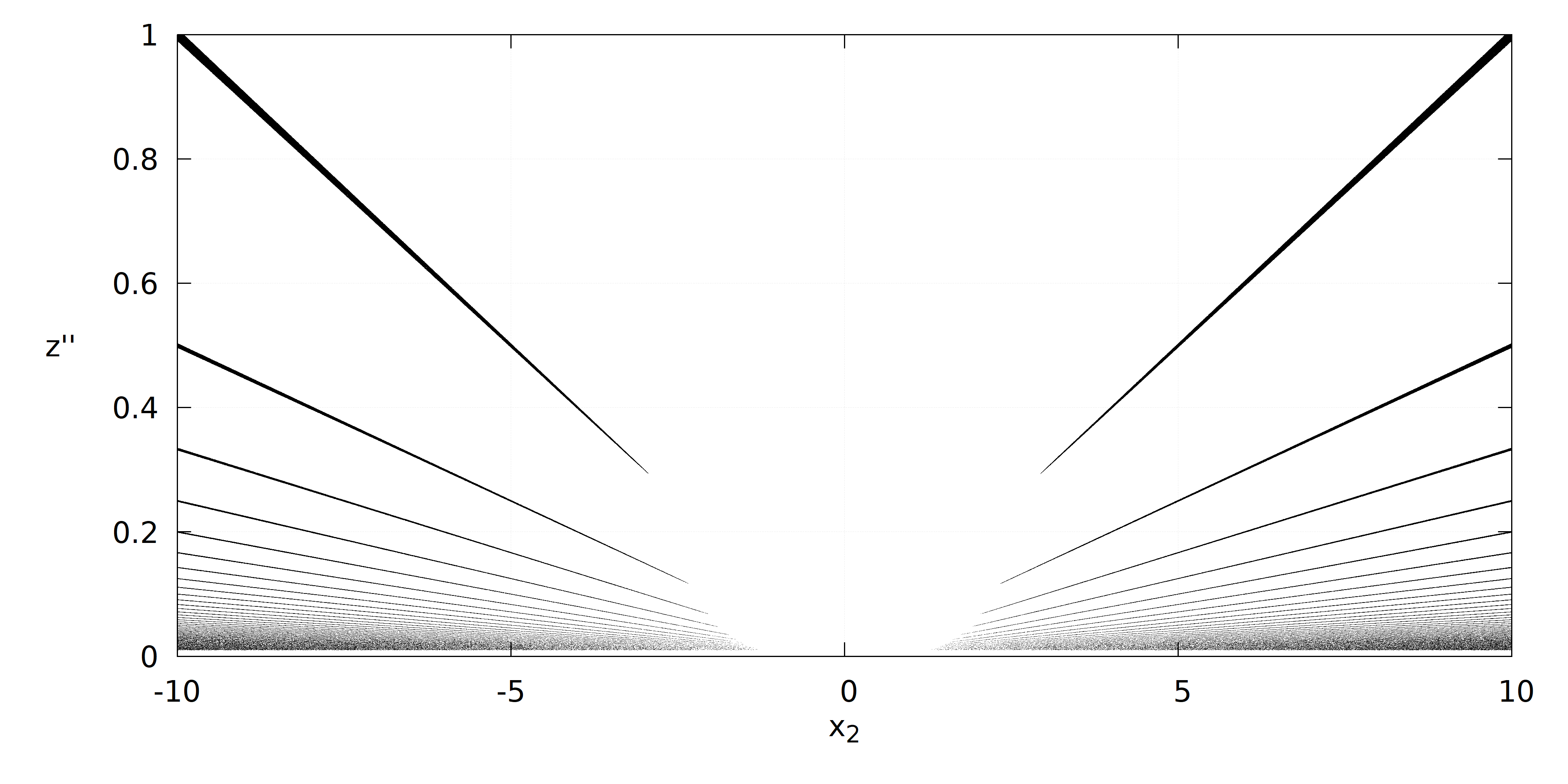}
    \caption{The Null map is obtained from the scattering through a single slit with a slit width of $0.1\lambda$. Null seems to be originating from a parabola like curve.}
    \label{fig: 1Slit SW0.1 }
\end{figure}

First, we investigate the Null maps obtained for the scattering through a single slit with a slit width of $0.1\lambda$, which is shown in Fig. \ref{fig: 1Slit SW0.1 }. The sub-$\lambda$ slit width is chosen to reduce the Gaussian points required to perform the numerical integration. However, the result will be identical to scaled slit-widths, if all the other parameters are scaled appropriately as was demonstrated explicitly in the subsection \ref{sec: scaling symmetry} (scaling symmetry).\\
As observed in Fig \ref{fig: 1Slit SW0.1 }, the Null maps have the diverging behavior. However, they do not form closed boundaries, which shows how the probability density can seep into the local maxima, which are surrounded by apparent Nulls. 
\begin{figure}
    \centering
    \includegraphics[width = 0.8\textwidth]{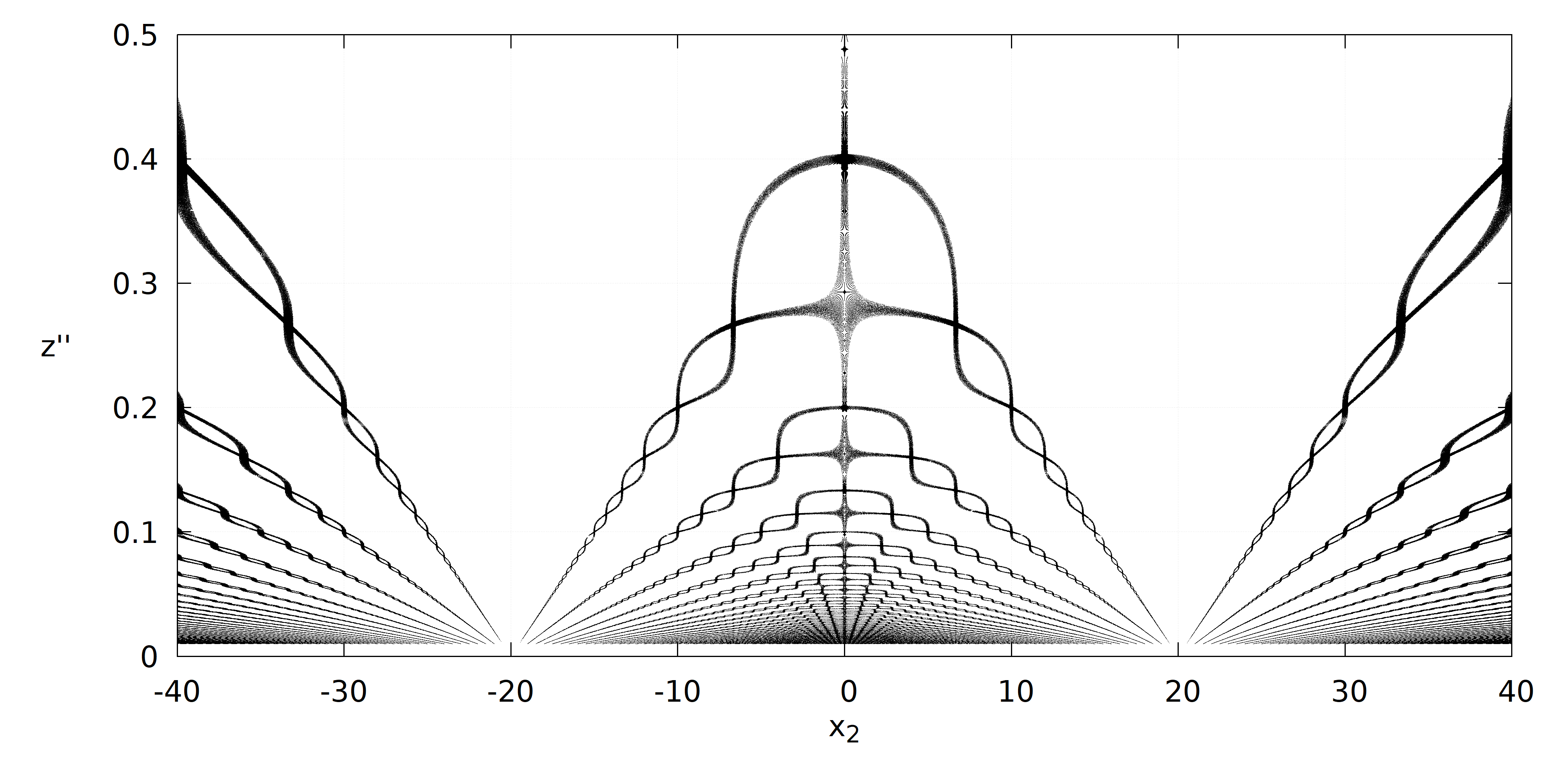}
    \caption{Null map of scattering through two slits of slit width $0.02\lambda$ and inter slit distance of $40\lambda$ has been shown. The Null appears to form braid like structure closer to the slit plane.}
    \label{fig: 2Slit SW 0.1 ISD 40}
\end{figure}
\begin{figure}
    \centering
    \includegraphics[width = 0.8\textwidth]{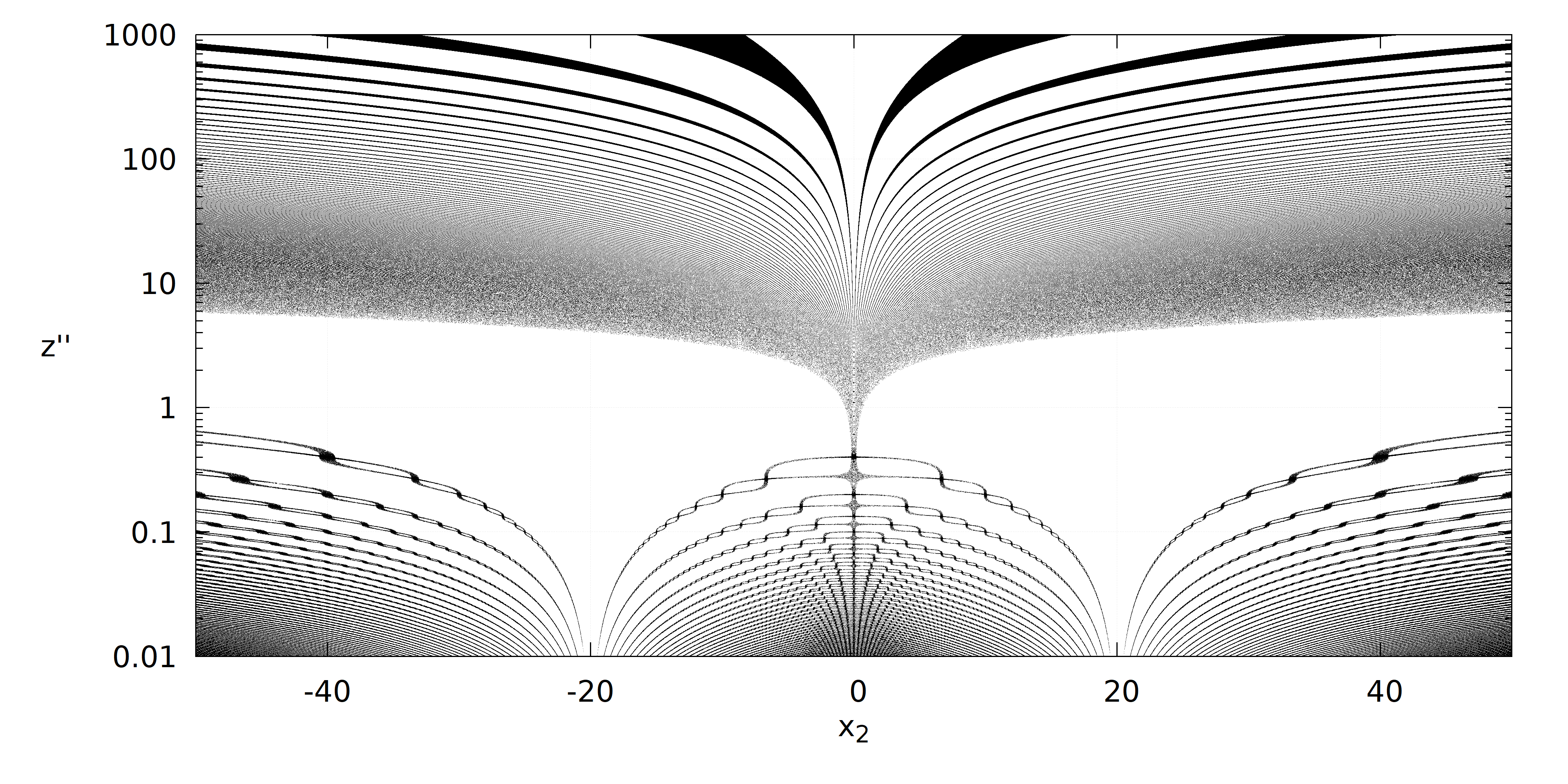}
    \caption{Null map plotted for same scenario as Fig. \ref{fig: 2Slit SW 0.1 ISD 40} but on log scale along $z^{\prime\prime}$. One notices a smooth transition occurring in the range $\approx 1$ to $\approx 8$ along $z^{\prime\prime}$.}
    \label{fig: 2 Slit SW 0.1 ISD 40 Log }
\end{figure}

The Null map of scattering through double-slit is demonstrated in the Fig. \ref{fig: 2Slit SW 0.1 ISD 40}, which reveals intriguing features, most notably are the braids like formation during the convolution of the nulls from the respective slits. At the current plotting scale the braids appear to be closed structures, but, it has been demonstrated in the previous subsection that having closed Null is essentially trapping a trajectory within it. However, in the scattering process, the trapped particle in what essentially looks like a bubble in the configuration space is counter-intuitive. Hence, it will require further investigation. We shall call the region enclosed by the Null boundary a Bubble. \footnote{One might notice Moir\'e patterns \cite{weisstein2002moire} in the Fig. \ref{fig: 2Slit SW 0.1 ISD 40} because of the overlaps of (a) limited density of the random point generator, (b) the underlying structure of Nulls, and (c) the chosen resolution of the plot.}
\begin{figure}
    \centering
    \includegraphics[width = 0.8\textwidth]{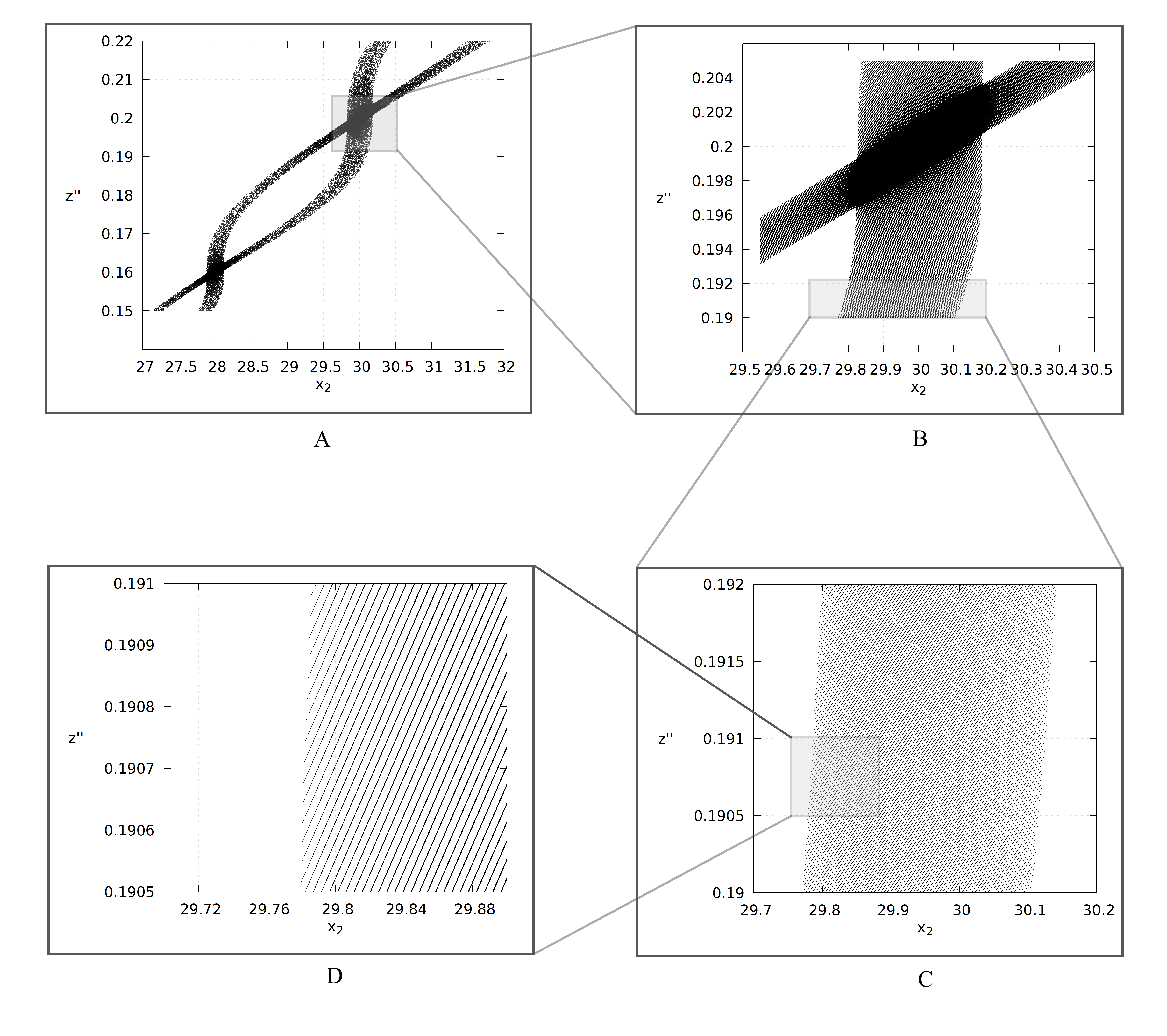}
    \caption{The zoom in of a braid of Fig. \ref{fig: 2Slit SW 0.1 ISD 40}. The braids which appeared to be closed are found to be perforated.}
    \label{fig: Braid Zoom }
\end{figure}

\begin{figure}
    \centering
    \includegraphics[width = 0.8\textwidth]{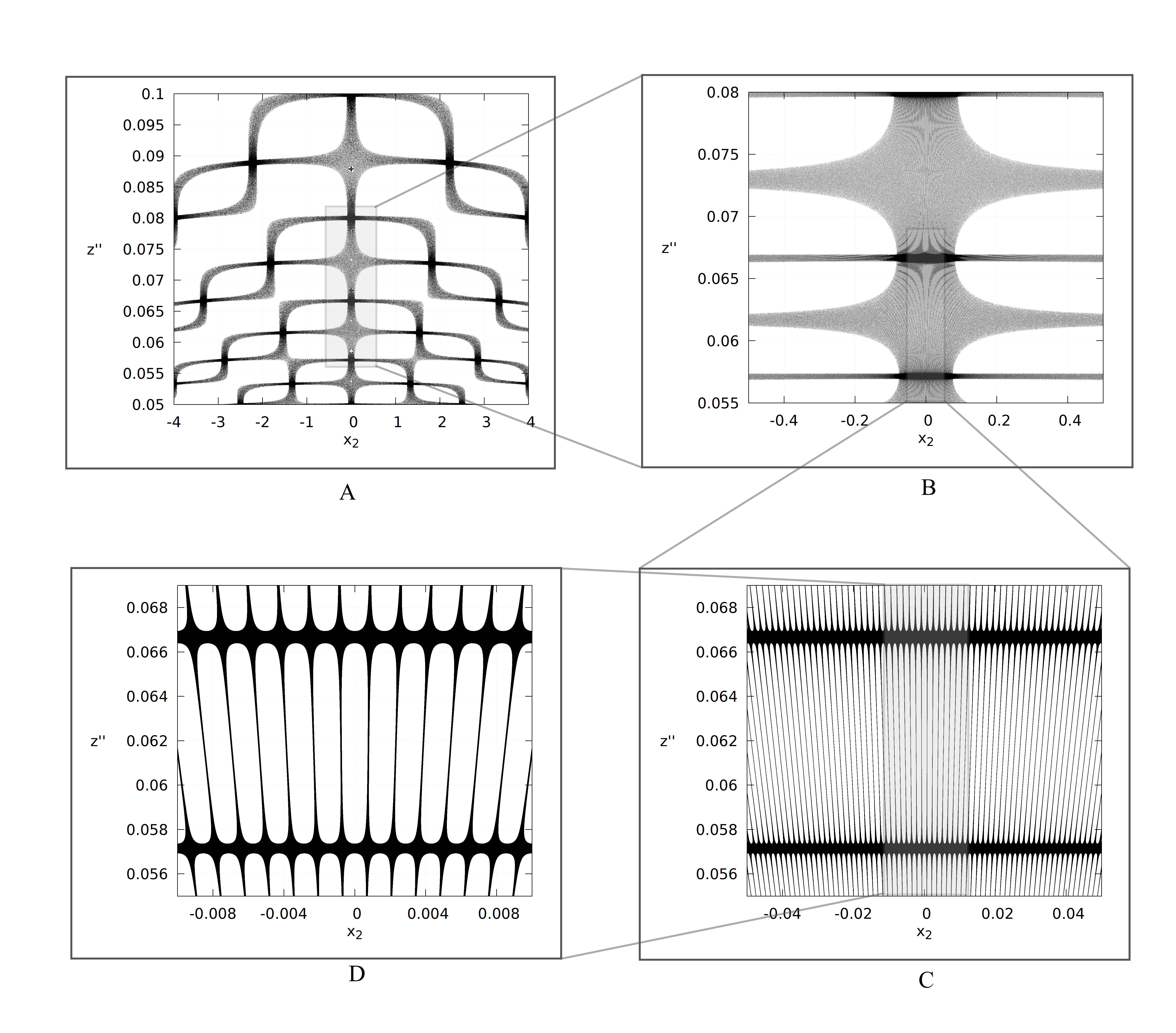}
    \caption{The zoom in the center of the Fig. \ref{fig: 2Slit SW 0.1 ISD 40}. An interesting structure is revealed in a region whose boundaries appear to be continuous closed nulls.}
    \label{fig: mid bubble Zoom }
\end{figure}
Upon zooming in the Fig. \ref{fig: 2Slit SW 0.1 ISD 40} we find that the braids form perforated boundaries (see Fig. \ref{fig: Braid Zoom }). However, a closer inspection of the region lying in the middle of the two slits reveals a Bubble like structure. It forms where the braids from the individual slit convolute with the ray of the Null originating from the center (Fig. \ref{fig: mid bubble Zoom }). 

\begin{figure*}
    \centering
    \begin{subfigure}{\textwidth}
        \includegraphics[width = 0.9\linewidth]{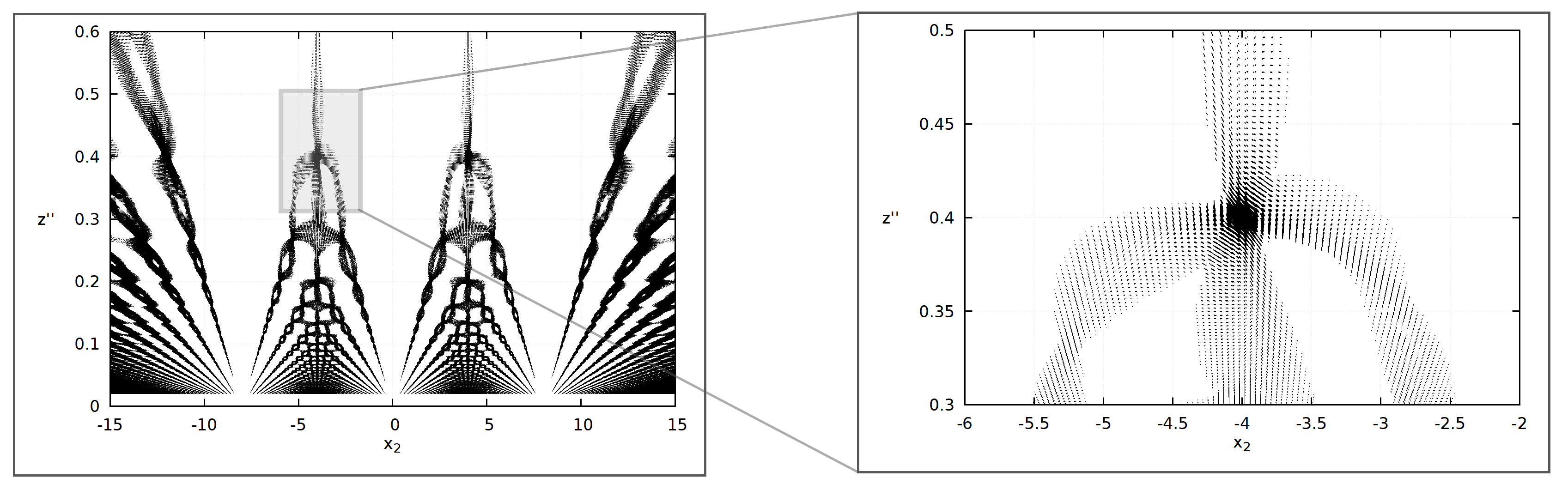}
        \vspace{0.5cm}
        \par\bigskip
        \caption{Null map of scattering through three slits.}
        \label{fig: 1D 3Slit NULL }
    \end{subfigure}
    \par\bigskip
    \par\bigskip
    \begin{subfigure}{\textwidth}
        \includegraphics[width = 0.89\linewidth]{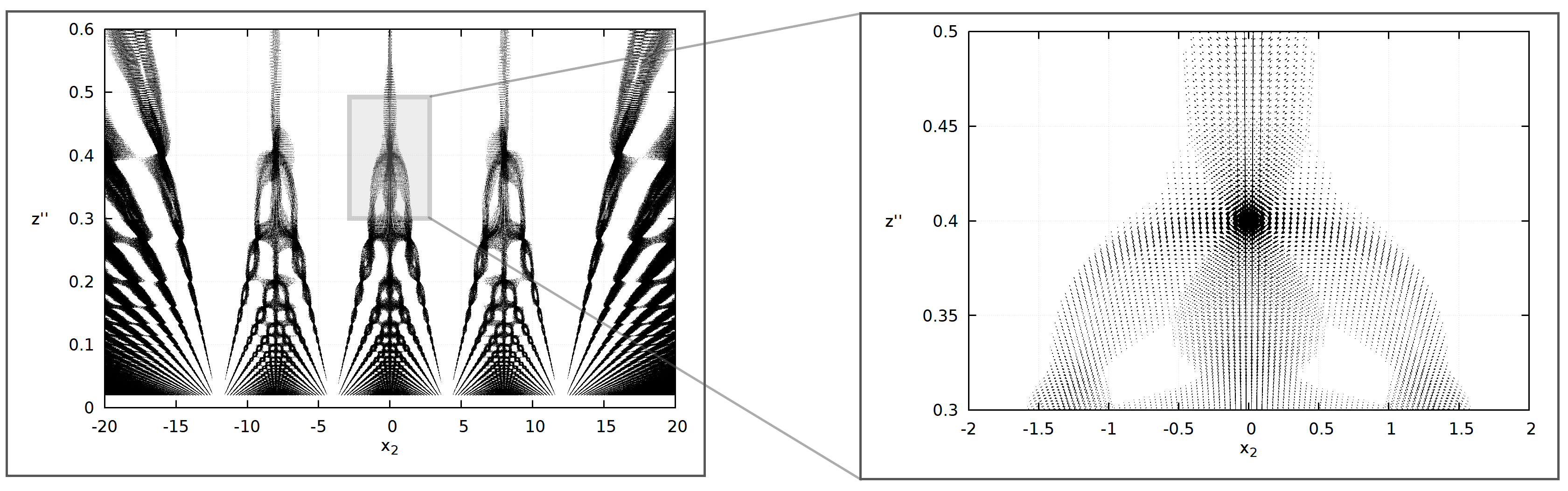}
        \vspace{0.5cm}
        \caption{Null map of scattering through four slits.}
        \label{fig: 1D 4Slit NULL }
    \end{subfigure}
    \par\bigskip
    \par\bigskip
    \begin{subfigure}{\textwidth}
        \includegraphics[width = 0.9\linewidth]{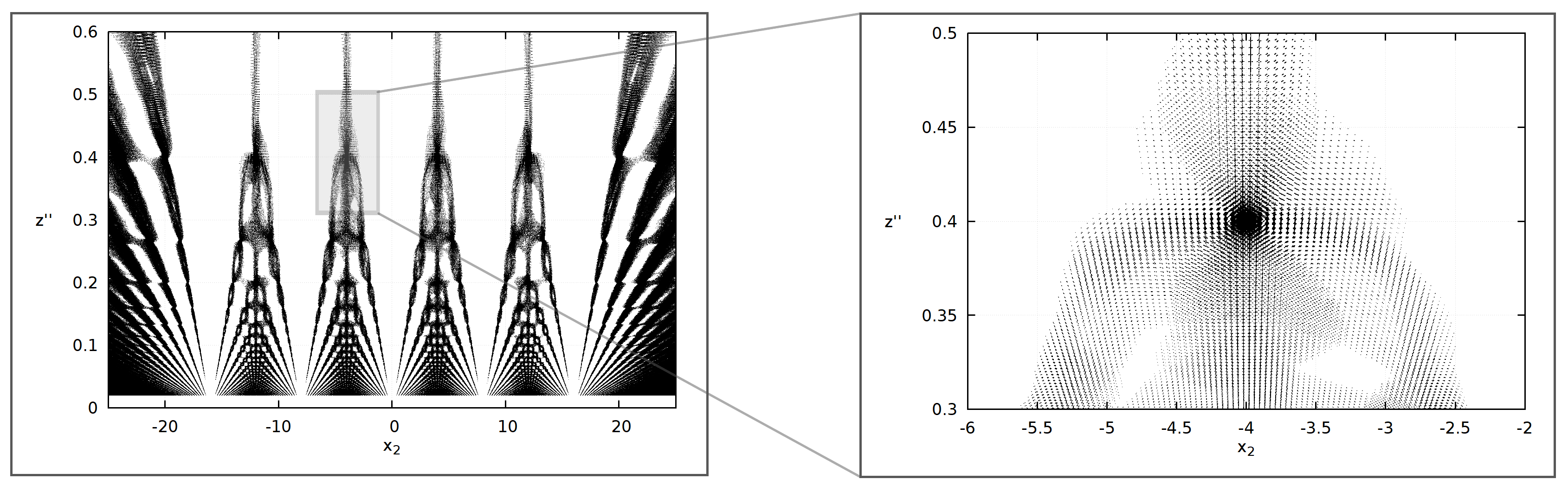}
        \vspace{0.5cm}
        \caption{Null map of scattering through five slits.}
        \label{fig: 1D 5Slit NULL }
    \end{subfigure}
    \par\bigskip
    \par\bigskip
    \caption{The above figures demonstrates increasing peculiarity for the increasing number of slits.}
    \label{fig: multi slit chaos}
\end{figure*}

The Null map far away from the slit plane undergoes a transition where the probability density from the individual slits merge together, which has been demonstrated in a log plot (Fig \ref{fig: 2 Slit SW 0.1 ISD 40 Log }), where transition is seen to be occurring somewhere in the interval $\approx 1$ to $\approx 8$ along $z^{\prime\prime}$. Analysis to obtain the transition has been done in the next section. Before proceeding into the analysis, the case of increasing the number of slits has been demonstrated in set of Figs. \ref{fig: multi slit chaos}. Where, one observes that adding slits makes the structure increasingly smeared and quite peculiar.

\section{Analysis}\label{sec: analysis}

\subsection{Hypergeometric Function}

\begin{figure*}
    \centering
    \begin{subfigure}{0.47\textwidth}
        \includegraphics[width = \textwidth]{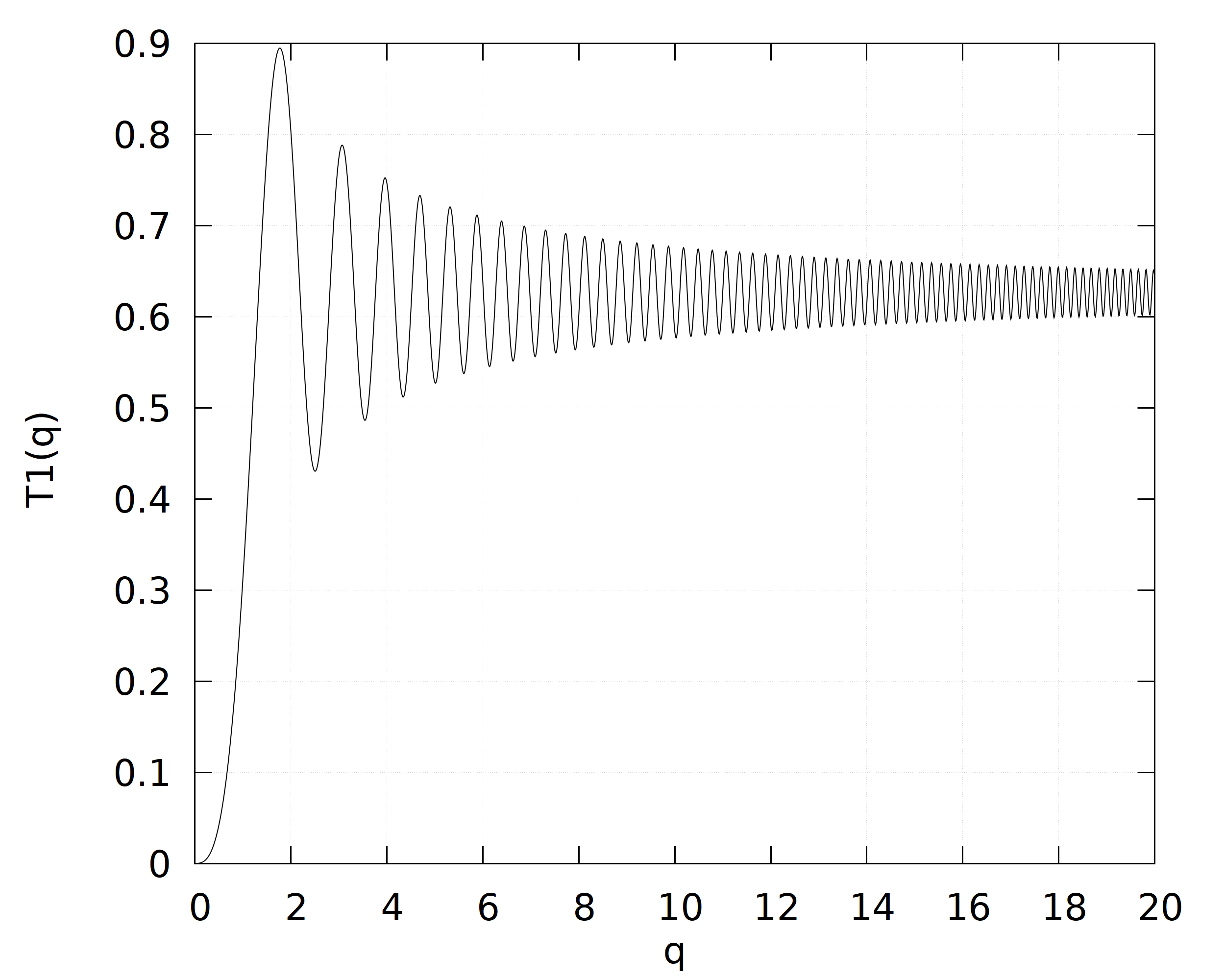}
        \caption{}
        \label{fig: T1 }
    \end{subfigure}
    \hfill
    \begin{subfigure}{0.47\textwidth}
        \includegraphics[width = \textwidth]{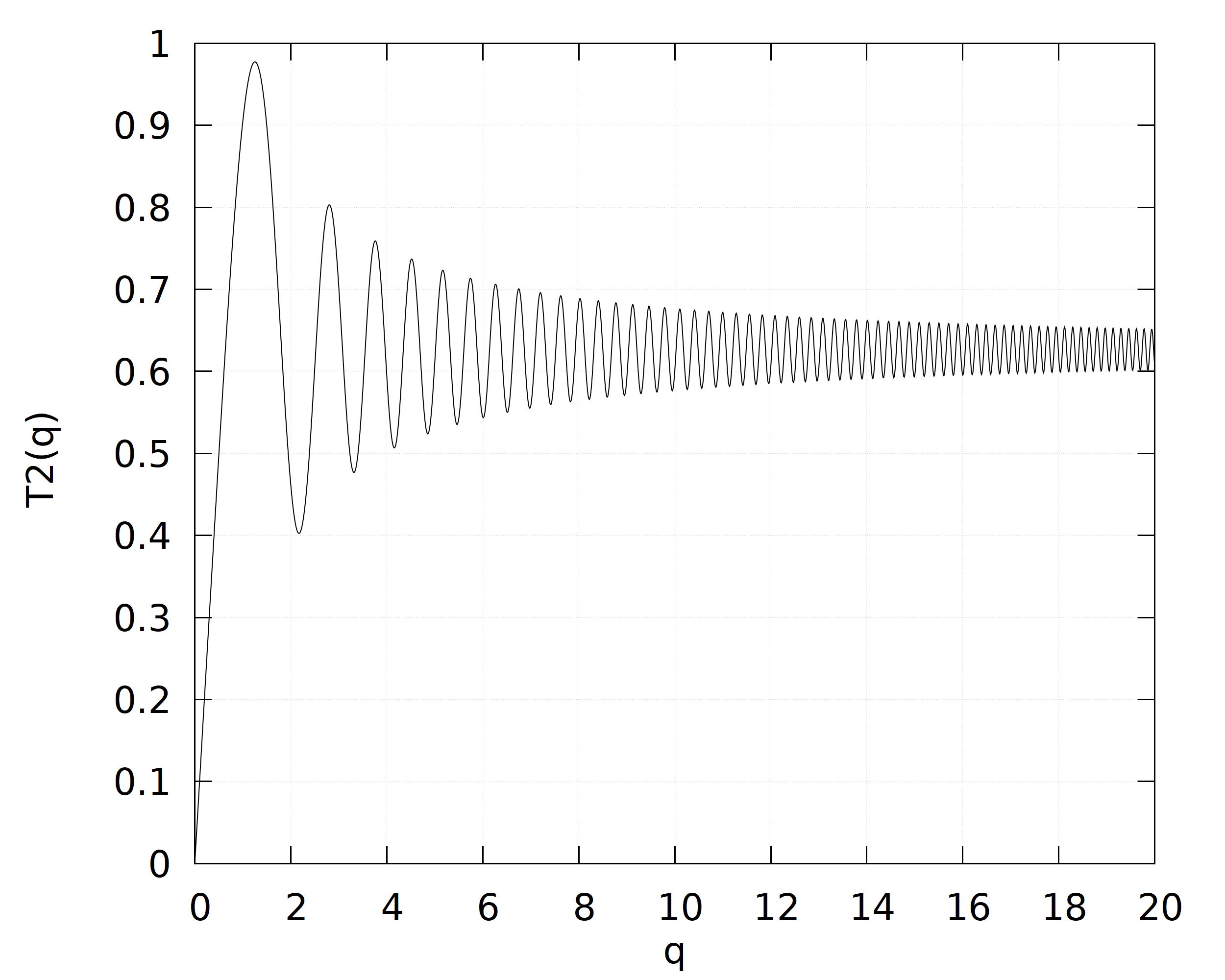}
        \caption{}
        \label{fig: T2 }
    \end{subfigure}
    
    \caption{$T_{1}(q)$ and $T_{2}(q)$ exhibit oscillatory behavior, with oscillation period getting shorter for large value of $q$}
    \label{fig: Hyp }
\end{figure*}
To understand the structures that have been obtained by numerical integration, we evaluate the integral by expanding it into an infinite series. Let us begin with Eqn. (\ref{eqn: 1d infinity source }). That is:
\begin{equation}
    \psi=\frac{1}{\sqrt{z^{\prime\prime}}}\int_{S^\prime} \exp{\iota\pi\frac{x^{\prime\prime2}}{z^{\prime\prime}}}\, dx_1
\end{equation}

Although the integral is pretty similar to the Gaussian integral, but it has complex argument, which makes it an error function. However, complex error functions are not easy to analyse. It is simpler to expand it into the Taylor series and conduct the analysis. Foremost, we make some simple substitutions. Using $q=\sqrt{\frac{\pi}{z^{\prime\prime}}}x^{\prime\prime}$, where $x^{\prime\prime}=x_1-x_2$,  $dx_1=\sqrt{\frac{z^{\prime\prime}}{\pi}}dq$. Note that as $x^{\prime\prime}\in\mathbb{R}$, which implies $q\ \in\mathbb{R}$.
Thus, the above integral becomes:
\begin{equation}
    \psi=\frac{1}{\sqrt\pi}\int_{S^\prime} \exp{\iota q^2}\, dq
\end{equation}
The set $S^\prime$ is appropriately scaled with the suggested substitution. Expanding the integrand in Taylor series:
\begin{equation}
    \psi=\frac{1}{\sqrt\pi}\int_{S^\prime}\sum_{k=0}^{\infty}\frac{\left(\iota q^2\right)^k}{k!}\, dq
\end{equation}
And, after evaluating the integral, one gets:
\begin{equation}
    \psi=\frac{1}{\sqrt\pi}\sum_{k=0}^{\infty}\frac{\iota^k}{k!}\frac{q^{2k+1}}{2k+1}\Biggm\vert_{ \partial S^\prime}
\end{equation}
Upon separating the real and imaginary parts, we get:
\begin{equation}\label{eqn: hyp_pre expanded }
    \psi=\frac{1}{\sqrt\pi}\Bigg[\iota\sum_{n=0}^{\infty}\left(-\right)^n\frac{q^{4n+3}}{\left(2n+1\right)!\,\left(4n+3\right)}
    +\sum_{n=0}^{\infty}\left(-\right)^n\frac{q^{4n+1}}{\left(2n\right)!\,\left(4n+1\right)} \Bigg] \Bigg\vert_{\partial S^\prime}
\end{equation}
Consider,
\begin{equation}\label{eqn: T1}
    T_1=\sum_{n=0}^{\infty}\left(-\right)^n\frac{q^{4n+3}}{\left(2n+1\right)!\,\left(4n+3\right)} 
\end{equation}
\begin{equation}\label{eqn:T2}
    T_2=\sum_{n=0}^{\infty}\left(-\right)^n\frac{q^{4n+1}}{\left(2n\right)!\,\left(4n+1\right)} 
\end{equation}
The above expression can be written in terms of hypergeometric series.
Foremost, we define shifted factorials as: $\forall\alpha\in\mathbb{C}\ $ and $\ k\geq0,\ k\ \in\mathbb{N}$
\begin{align}
    \left(\alpha\right)_k & =\prod^{k-1}_{j=0} (\alpha + j)\\
    \left(\alpha\right)_0 & =1
\end{align}
By this definition, it is easy to see that:
\begin{equation}\label{eqn: 4n+3 }
    \frac{4n+3}{3}=\frac{\left(7/4\right)_n}{\left(3/4\right)_n}
\end{equation}
Identifying:
\begin{gather}
    \left(2n+1\right)!=4^n\left(3/2\right)_n n!\\ \left(4n+3\right)=\frac{\left(7/4\right)_n}{\left(1/3\right)\left(3/4\right)_n}
\end{gather}
Substituting the above in Eqn: \ref{eqn: T1} gives us:
\begin{equation}
    T_1=\frac{q^3}{3}\sum_{n=0}^{\infty}\frac{\left(3/4\right)_n\left(-q^4/4\right)^n}{n!\left(3/2\right)_n\left(7/4\right)_n}
\end{equation}
The above sum is a Hypergeometric series:
\begin{equation}
    {}_1F_2\left(3/4;3/2,7/4;-q^4/4\right)=\sum_{n=0}^{\infty}\frac{\left(3/4\right)_n\left(-q^4/4\right)^n}{n!\left(3/2\right)_n\left(7/4\right)_n}
\end{equation}
This series is convergent for all $q \in \mathbb{C}$ \cite{andrews1999special}. Thus $T_1$ in terms of Hypergeometric series is written as:
\begin{equation}
    T_1=\frac{q^3}{3} {}_1F_2\left(3/4;3/2,7/4;-q^4/4\right)
\end{equation}
By similar procedures, $T_2$ can also be expressed as:
\begin{equation}
    T_2=q\,{}_1F_2\left(1/4;1/2,5/4;-q^4/4\right)
\end{equation}
Therefore:
\begin{equation}
    \psi=\frac{1}{\sqrt\pi}\Bigg[q\,{}_1F_2\left(1/4;1/2,5/4;-q^4/4\right)
    +\iota\frac{q^3}{3}{}_1F_2\left(3/4;3/2,7/4;-q^4/4\right)\Bigg]\Bigg\vert_{\partial S^\prime}
\end{equation}

The hypergeometric functions obtained above are oscillatory for argument less than zero.
Consider the case of two slits for simplicity. We assume that the slits are symmetric with respect to the point $x_1=0$. Then the interval over which integration is carried out can be expressed as $x_1\in\left[-\beta,-\alpha\right]\cup\left[\alpha,\beta\right]$, here, $-\beta$, $-\alpha$, $\alpha$ and $\beta$ are the edges of the slits. Correspondingly the integration limits are then expressed as: 
\begin{gather}
    q_1=\sqrt{\frac{\pi}{z^{\prime\prime}}}\left(-\beta-x_2\right)\\
    q_2=\sqrt{\frac{\pi}{z^{\prime\prime}}}\left(-\alpha-x_2\right)\\
    q_3=\sqrt{\frac{\pi}{z^{\prime\prime}}}\left(\alpha-x_2\right)\\
    q_4=\sqrt{\frac{\pi}{z^{\prime\prime}}}\left(\beta-x_2\right)
\end{gather}

With the hypergeometric analysis, we can also convince ourselves how the Null map is exhibiting peculiarity with increasing number of slits. Let $\alpha_i$ be an edge of a given slit. Hence, $q_i=\sqrt{\frac{\pi}{z^{\prime\prime}}}\left(\alpha_i-x_2\right)$. Therefore, we can express the wave-function (From Eqn. (\ref{eqn: hyp_pre expanded })):
\begin{equation}
    \psi=\frac{1}{\sqrt\pi}\sum_{i}{\left(-\right)^i\left[T_1\left(q_i\right)+\iota T_2\left(q_i\right)\right]}
\end{equation}

Let us consider a path, such that one of the $q_i$ is a constant on it. Let us call the constant $q_k$. Thus one can write:
\begin{equation}\label{eqn: q const path}
    x_2=\alpha_k-\frac{q_k}{\sqrt\pi}\sqrt{z^{\prime\prime}}
\end{equation}
Upon substituting $x_2$ thus obtained in the other $q_i$'s we get:
\begin{equation}
    q_i=\sqrt{\frac{\pi}{z^{\prime\prime}}}\left(\alpha_i-\alpha_k\right)+q_k
\end{equation}
For the case of single slit, the $i$ index has only two values. Upon fixing $q_k$, the other $q_i$ varies along the path (Eqn. (\ref{eqn: q const path})) to contribute to the fluctuations. Hence, we observe fairly simple oscillations. It is also interesting to note that the Eqn. (\ref{eqn: q const path}) represent a parabolic path, which can be traced near the origin in Fig. \ref{fig: 1Slit SW0.1 }. However, upon adding a second slit ($i$ index runs from 1 to 4), and fixing one $q_k$ we are left with three variable $q_i$s. However, it is not straightforward how having more variable $q_i$s leads to the peculiar behaviour. This can be understood if we consider that the slit-width is much smaller than inter-slit distance. This consideration implies that the contribution from any individual slit comes from a particular region of $T_1$ and $T_2$ functions. A closer inspection of Fig. \ref{fig: Hyp } reveals that the period of oscillation of $T_1$ and $T_2$ functions varies with $q$. It is known that the superposition of different periods leads to beats; this suggests the reason for observing such a quasi-periodic fluctuation in the probability density as found in Fig. \ref{fig: 2Slit SW 0.1 ISD 40}. Furthermore, if we include a third slit, we will have a contribution from another set of points, thus, adding another period onto the chosen path. This is what appears to be happening in Fig. \ref{fig: multi slit chaos}, which seems to account for the peculiar behaviour that is reported here.

\subsection{Fresnel Function}\label{fresnel}
\begin{figure*}
    \centering
    \begin{subfigure}{0.47\textwidth}
        \includegraphics[width = \linewidth]{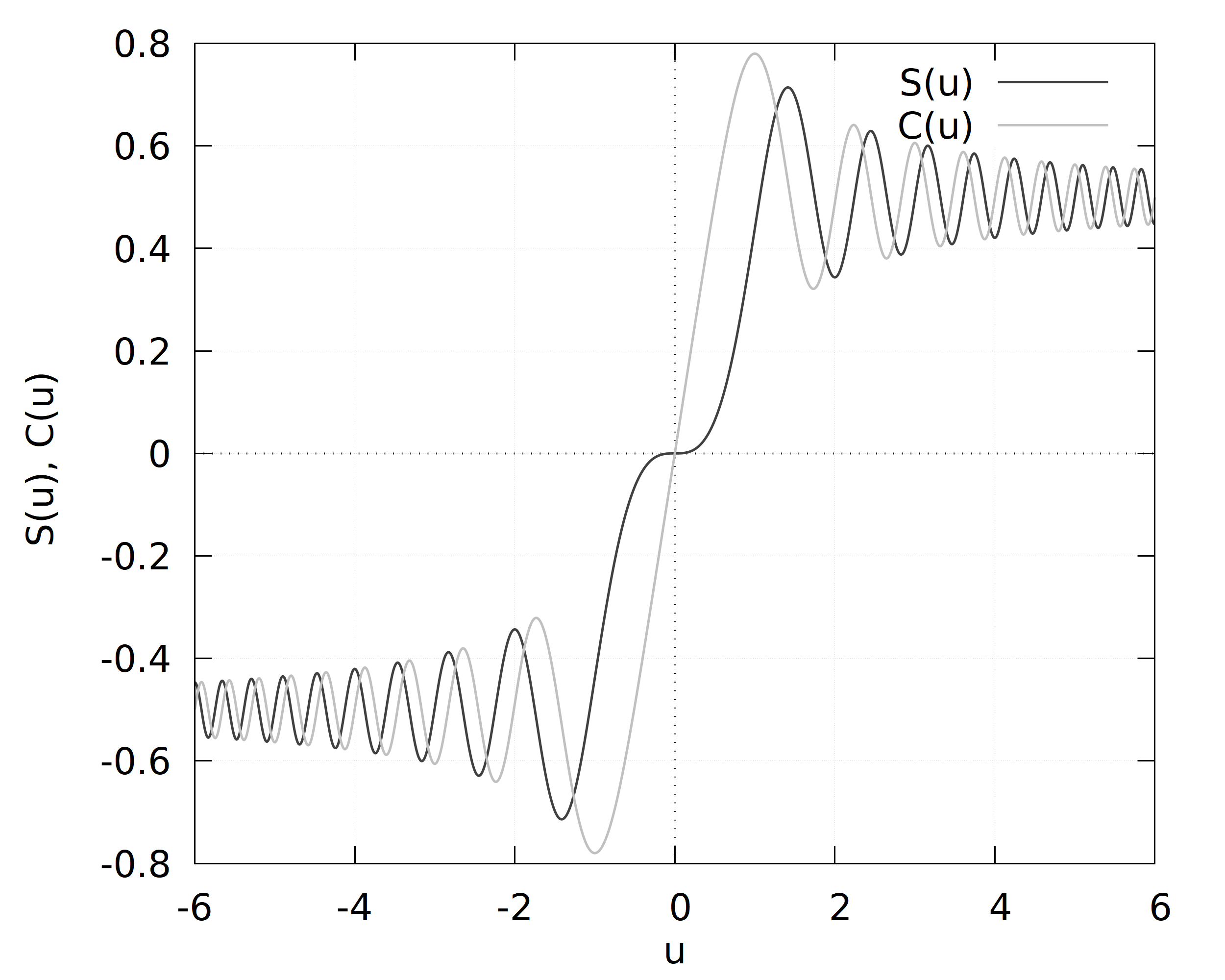}
        \caption{}
        \label{fig: Fresnel}
    \end{subfigure}
    \hfill
    \begin{subfigure}{0.47\textwidth}
        \includegraphics[width = \linewidth]{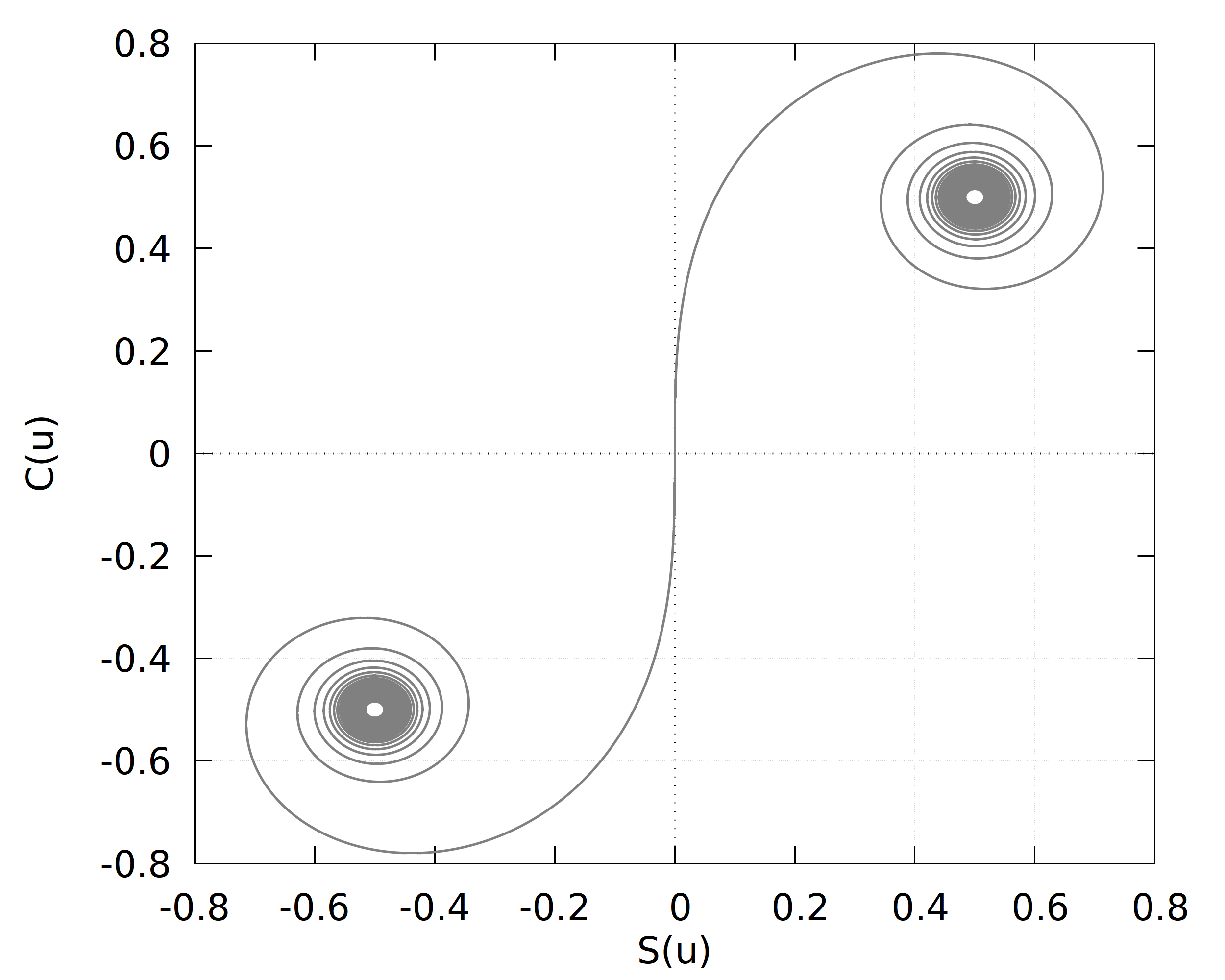}
        \caption{}
        \label{fig: Cornu}
    \end{subfigure}
    \caption{The Fresnel plot is shown in Fig. A; both are odd functions and exhibit oscillatory behavior with an apparent phase difference. The parametric plot of the Fresnel functions is shown in Fig. B, which is known as Cornu spiral, also known as the Euler Spiral.}
    \label{fig: Fresnel and cornu}
\end{figure*}

The parametric plot obtained by plotting $T_1$ and $T_2$ yields a Cornu like spiral. The Cornu spiral has been applied extensively in the theory of Fresnel diffraction (see, for example,\cite{born2013principles}). With this motivation, we next analyse the patterns obtained here by directly using the Cornu spiral and its properties.\\
Conventionally the Cornu spiral is a parametric plot generated by using the Fresnel functions, which are defined as:
\begin{gather}
    S\left(z\right)=\int_{0}^{z}sin\left(\pi u^2/2\right)\, du\\
    C\left(z\right)=\int_{0}^{z}cos\left(\pi u^2/2\right)\, du\\
    S\left(z\right)+\iota C\left(z\right)=\int_{0}^{z}\exp{\iota\pi u^2/2}\, du \label{eqn: Fresnel functions }
\end{gather}
These functions are identical to $T_1$ and $T_2$ up to an overall multiplicative factor. Therefore the conclusions drawn using the Cornu spiral are applicable to the case under consideration.\\
In terms of the Fresnel functions the wave-function, up to an overall factor of $1/\sqrt{2}$, can be expressed as: 
\begin{equation}
    \psi=\left[S\left(u\right)+\iota C\left(u\right)\right] \Big\vert_{\partial S^\prime}
\end{equation}
Expressing the wave-function in the form of Fresnel functions presents an interesting way of analysis, as it enables us to comment on its behavior while looking at the Cornu spiral directly. The Fresnel functions along with Cornu spiral for ready reference have been shown in Fig. \ref{fig: Fresnel and cornu}.

\begin{figure*}
    \centering
    \begin{subfigure}{\textwidth}
        \includegraphics[width = 0.95\linewidth]{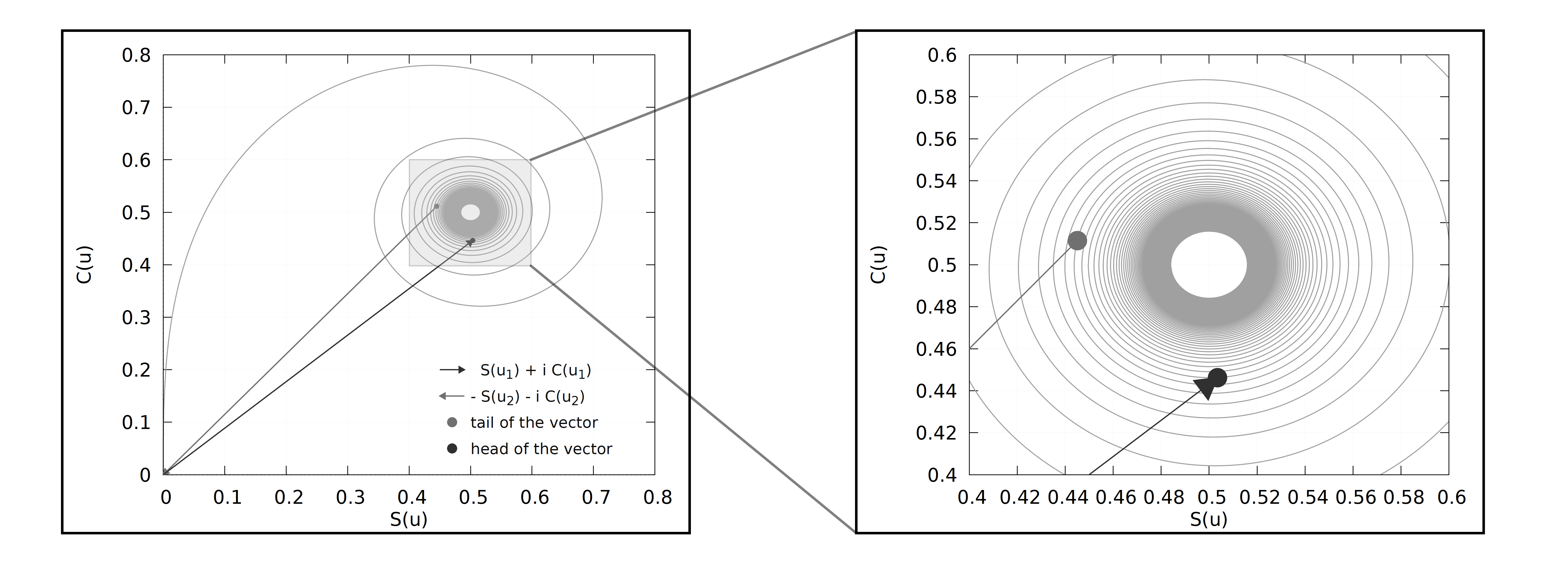}
        \caption{Sum of two vectors onto the Cornu spiral will represent the Probability amplitude at a given point after getting scattered through a single slit.}
        \label{fig: cornu 1slit 1 }
    \end{subfigure}
    
    \begin{subfigure}{\textwidth}
        \includegraphics[width = 0.95\linewidth]{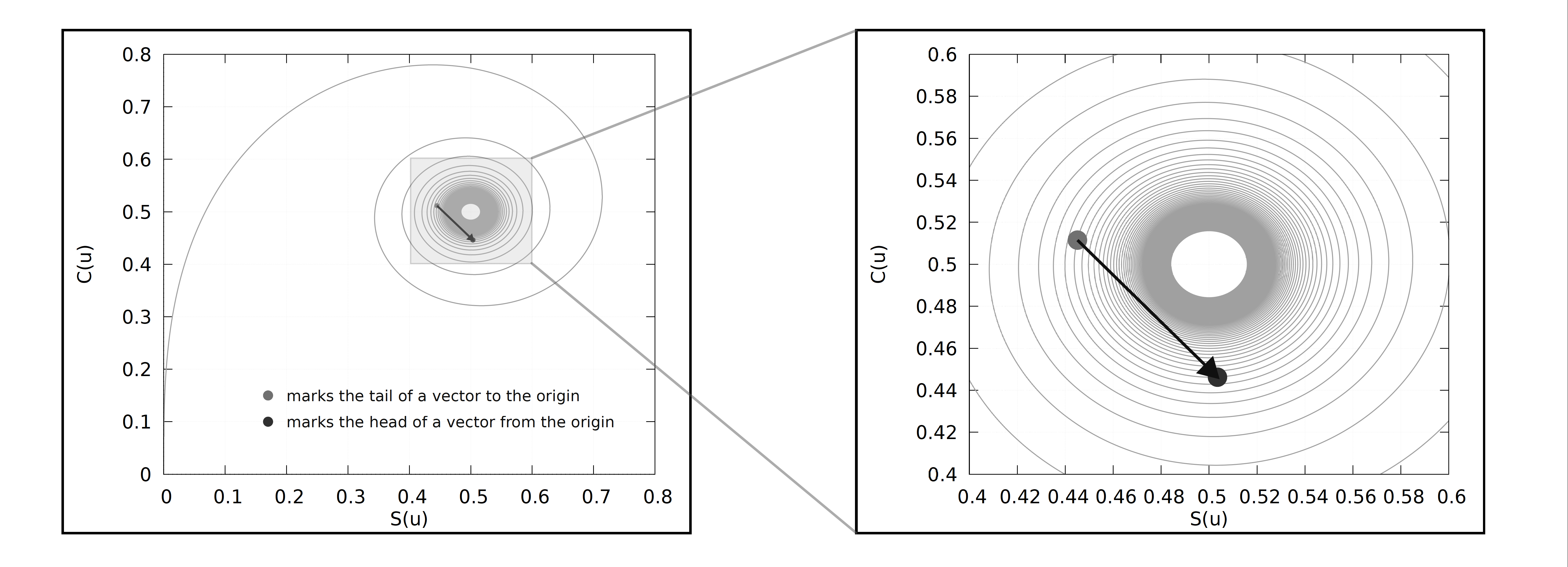}
        \caption{Vectors from origin are represented by only dots for brevity. The Probability amplitude at a given point will be determined by the vector drawn between the two dots as shown above.}
        \label{fig: Cornu 1slit 2 point}
    \end{subfigure}
    
    \begin{subfigure}{\textwidth}
        \includegraphics[width = 0.95\linewidth]{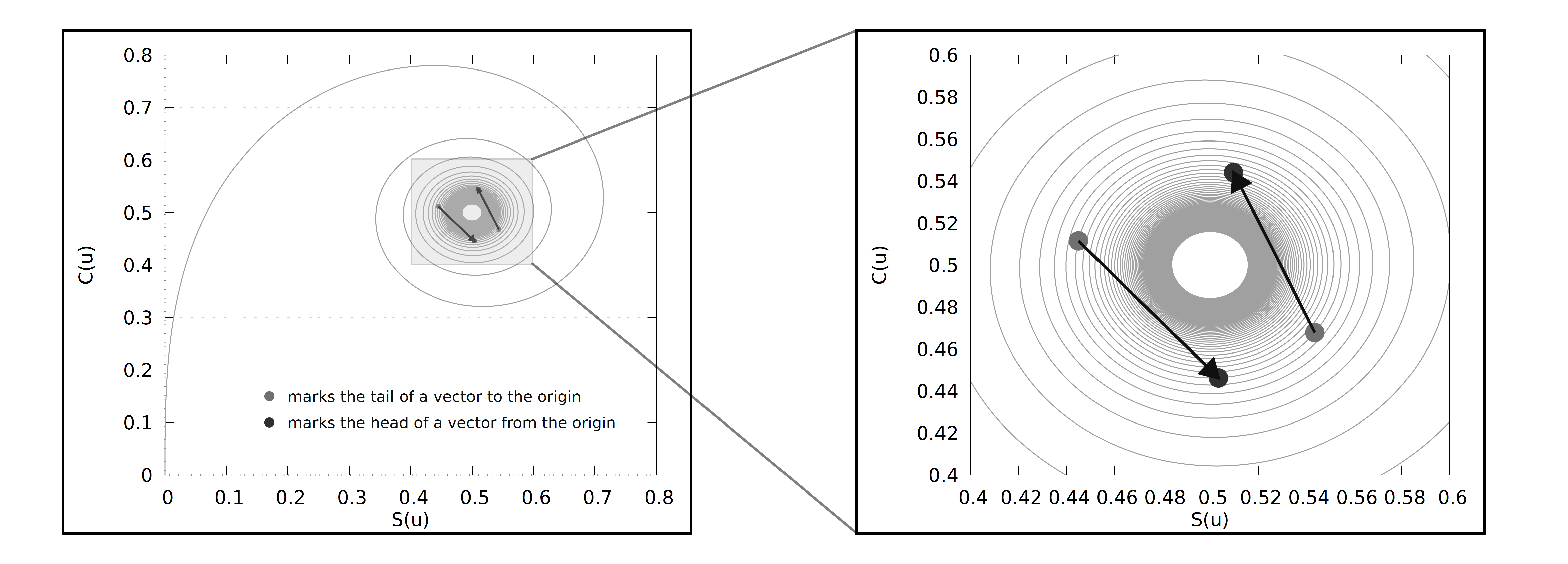}
        \caption{In the case of double-slit, there will be four vector dots on the Cornu spiral. The difference in the vectors shown above will determine the probability amplitude at a given point.}
        \label{fig: Cornu 2slit 1 point}
    \end{subfigure}

    \caption{Vector representation onto the cornu spiral has been shown in the above figures, and a method of estimating the probability amplitude has been demonstrated. }
    \label{fig: vectors onto cornu}
\end{figure*}

\begin{figure}
    \centering
    \includegraphics[width = 0.8\textwidth]{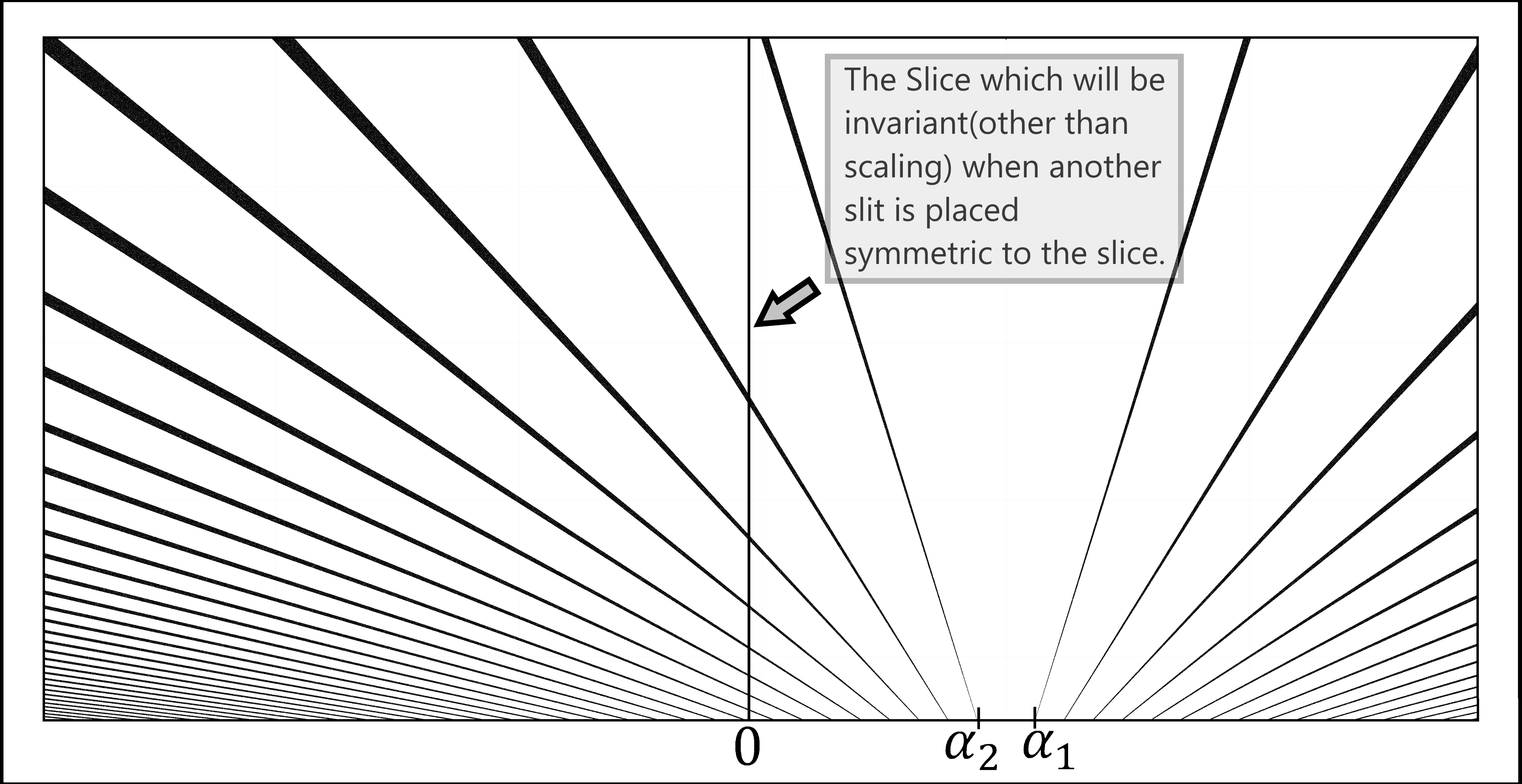}
    \caption{The Null map of the single slit whose edges are at $\alpha_1$ \& $\alpha_2$ is illustrated above, if another slit is introduced at $-\alpha_1$ \& $-\alpha_2$, then the marked slice in the center will get scaled. However, the relative behavior along the slice will remain unchanged as it was in the presence of single slit.}
    \label{fig: single slit slice }
\end{figure}

\begin{figure}
    \centering
    \includegraphics[width = 0.8\textwidth]{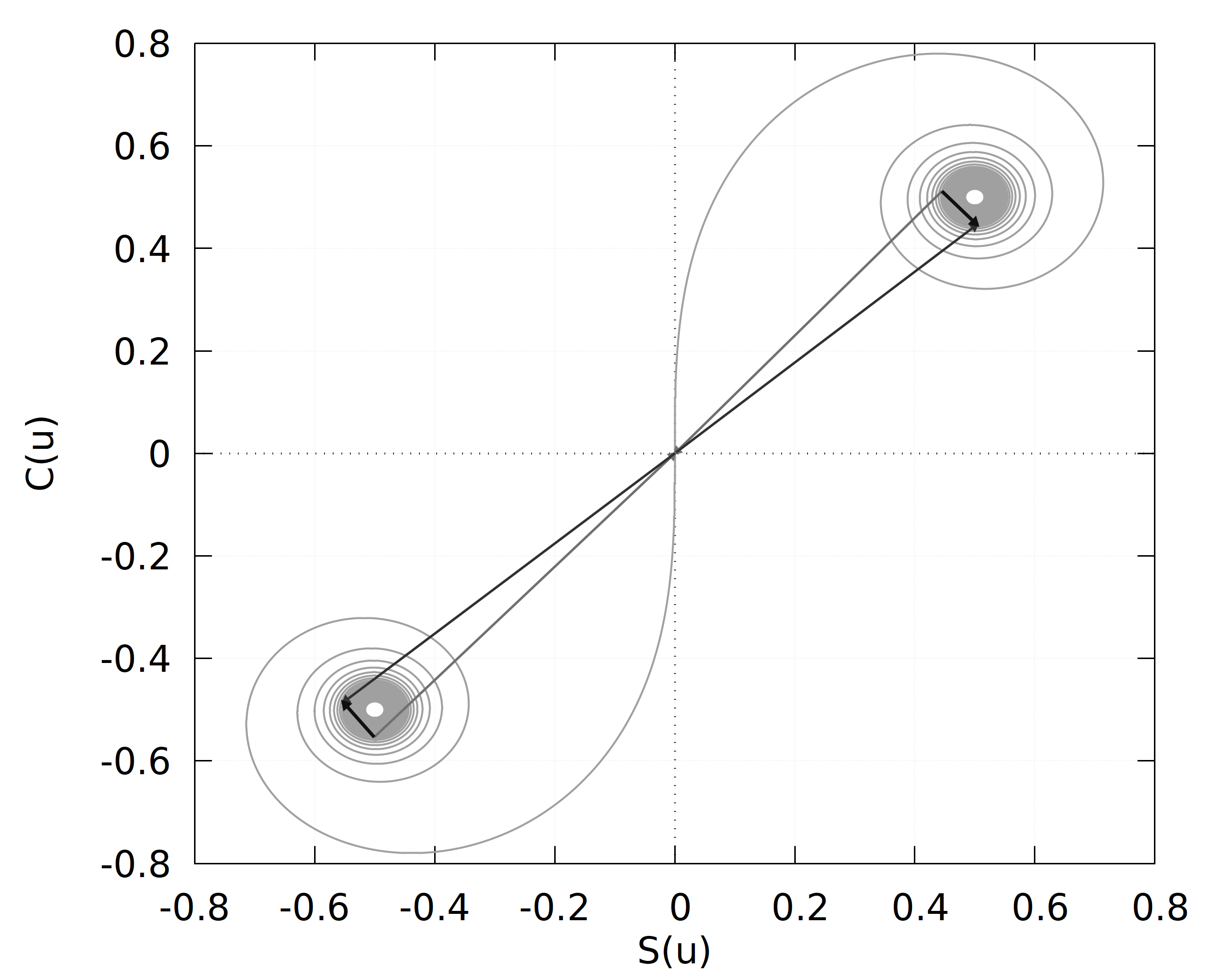}
    \caption{An illustration of the arrangement of vectors near the center of the viewing plane has been shown above. As they lie in the opposite quadrants, a slight shift from the center can move the difference vector facing opposite to each other, thus creating a null. A path conserving this relative orientation will form a restrictive boundary for the trajectory.}
    \label{fig: cornu mid null}
\end{figure}

Given $u_i=\sqrt2\left(\alpha_i-x_2\right)/\sqrt{z^{\prime\prime}}$, the wave-function is written as:
\begin{equation}
    \psi=\sum_{i}\left(-\right)^{i+1}\left[S\left(u_i\right)+\iota C\left(u_i\right)\right]
\end{equation}
A bit of reflection reveals that $\left[S\left(u_i\right)+\iota C\left(u_i\right)\right]$ can be represented as a vector from origin to a point on the Cornu spiral. This representation will imply that $\psi$ is the sum of vectors on the parametric plot with alternating signs. 
In order for $\psi$ to be a Null, these vectors should sum up to zero. 

In the case of a single slit, we have only two vectors with opposite signs. For them to form a Null, they are required to be identical. However, every $u_i$ corresponds to a unique point on the Cornu spiral. Given that the Cornu spiral has no self intersections, it follows that these vectors never add to form a Null. Hence, we conclude that there is no true Null in the case of the single slit. Thus, in Fig. \ref{fig: 1Slit SW0.1 }, the Nulls maps are not made of true Nulls but they represent ponts at which the probability density is smaller than $10^{-14}$ but non-zero.

In the case of the double-slit, we are dealing with four vectors, whose sum has to lead to a Null. We can divide them into two sets, each set containing vectors from each slit. Thus, to have a net Null, vector sum of both sets must yield equal vectors, but with opposite signs, which is not forbidden on the parametric plot. Hence, $\psi$ can be zero at some points in the case of the double-slit (see Fig. \ref{fig: Cornu 2slit 1 point} for illustration). Thus, Null points mapped in Fig. \ref{fig: 2Slit SW 0.1 ISD 40} can be actual Nulls.\\
Now, we begin the analysis of the bubble found in Fig. \ref{fig: mid bubble Zoom }. First, we explore the exact nature of the bubble at the central slice described by $x_2=0$. This simplifies $u_i$, i.e., $u_i=\sqrt{2/z^{\prime\prime}}\alpha_i$. Since, the $\alpha_i$ are symmetric with respect to the origin, one obtains, $\alpha_1=-\alpha_4$ and $\alpha_2=-\alpha_3$. Noting further that Fresnel integrals are odd functions, one gets:\\
\begin{gather}
    S\left(u_4\right)+\iota C\left(u_4\right)=-\left(S\left(u_1\right)+\iota C\left(u_1\right)\right)\\
    S\left(u_3\right)+\iota C\left(u_3\right)=-\left(S\left(u_2\right)+\iota C\left(u_2\right)\right)
\end{gather}
Therefore the wave-function along the slice $x_2=0$ is:
\begin{equation}
    \psi=2\left[\left(S\left(u_1\right)+\iota C\left(u_1\right)\right)-\left(S\left(u_2\right)+\iota C\left(u_2\right)\right)\right]
\end{equation}
which is identical to the wave-function of single-slit taken along the same slice (illustrated in Fig. \ref{fig: single slit slice }). However, it has been demonstrated that no Null exists in the case of a single slit; thus, the horizontal boundary that we encounter on this path is not true Null. Hence, establishing that the bubble observed in Fig. \ref{fig: mid bubble Zoom } is not completely enclosed, and the probability density flow is permitted through the upper and lower edges of the bubble. We shall call these structures quasi-bubbles for the sake of brevity. Since, $\rho$ is a continuous function, having a point at which $\rho$ is non-zero guarantees that it is non-zero over a finite region.\\
However for sides (Fig. \ref{fig: mid bubble Zoom } D), it can be argued that there can exist a true Null, as demonstrated below.\\ 
\begin{figure*}[ht]
    \centering
    \begin{subfigure}{0.62\textwidth}
        \includegraphics[width = \linewidth]{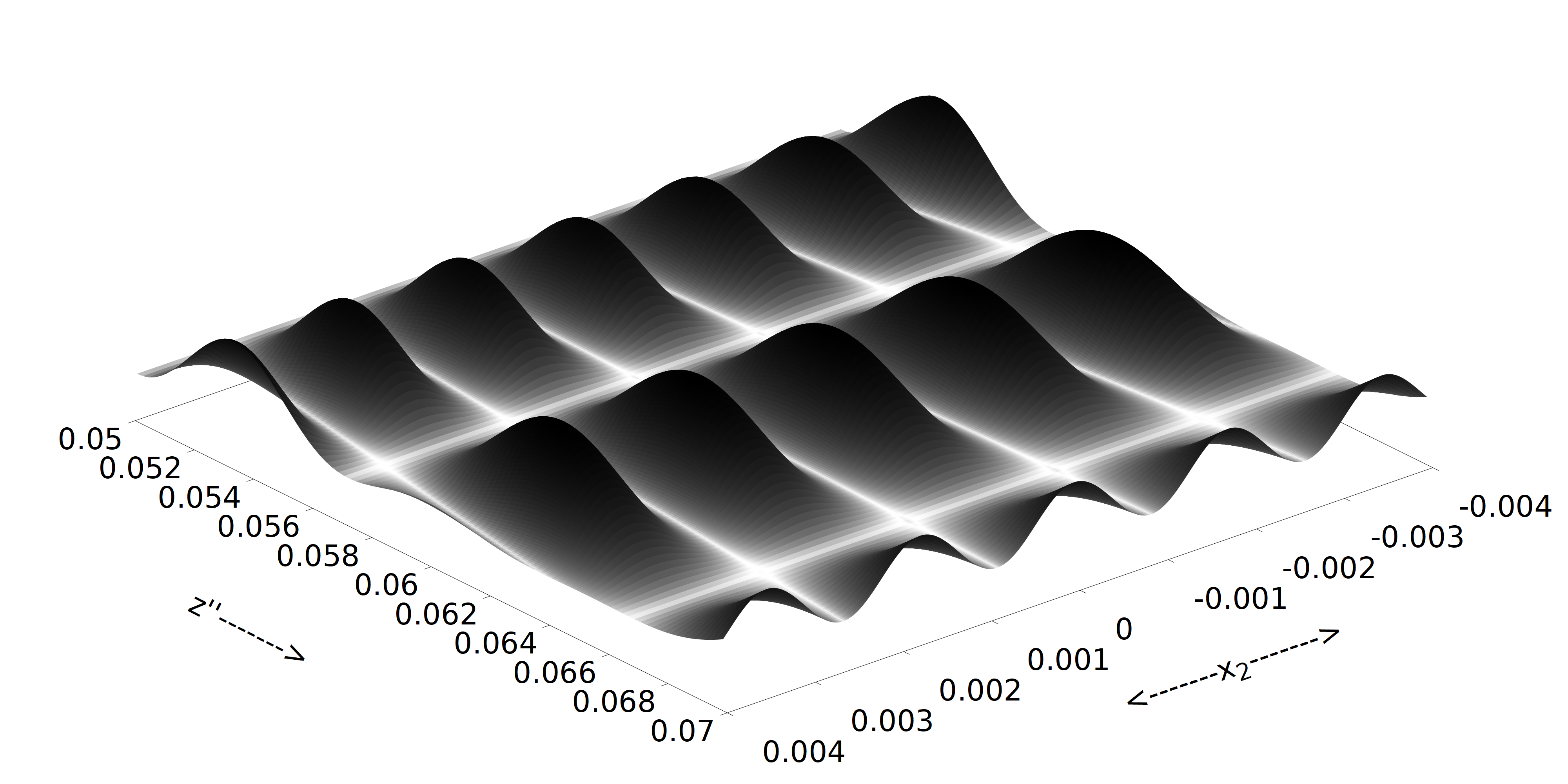}
        \caption{}
        \label{fig: bubble heat}
    \end{subfigure}
    \hfill
    \begin{subfigure}{0.37\textwidth}
        \includegraphics[width = \linewidth]{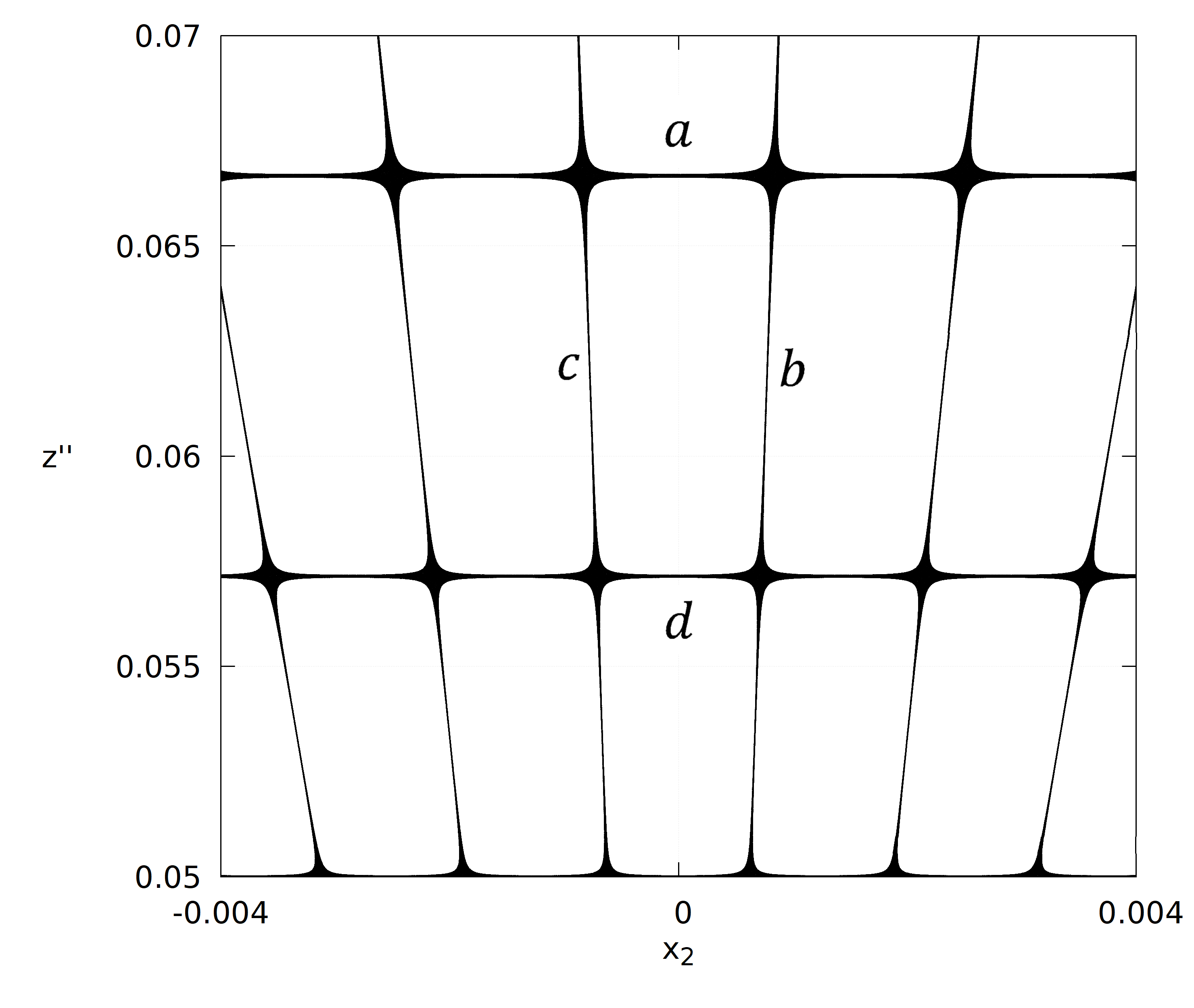}
        \caption{}
        \label{fig: bubble heat null}
    \end{subfigure}
    \caption{Fig. (A) shows the bubble as has been obtained from the numerical integration. Fig. (B) shows the Null map of the same region. The bubble appears to be closed. The integration is exact for 1-D, where the $z^{\prime\prime}$ is the temporal direction. Thus, one observes the change in probability density with time which is enclosed between the Null.}
    \label{fig: bubble and continuity}
\end{figure*}

\begin{figure*}
    \centering
    \begin{subfigure}{0.49\textwidth}
        \includegraphics[width = \linewidth]{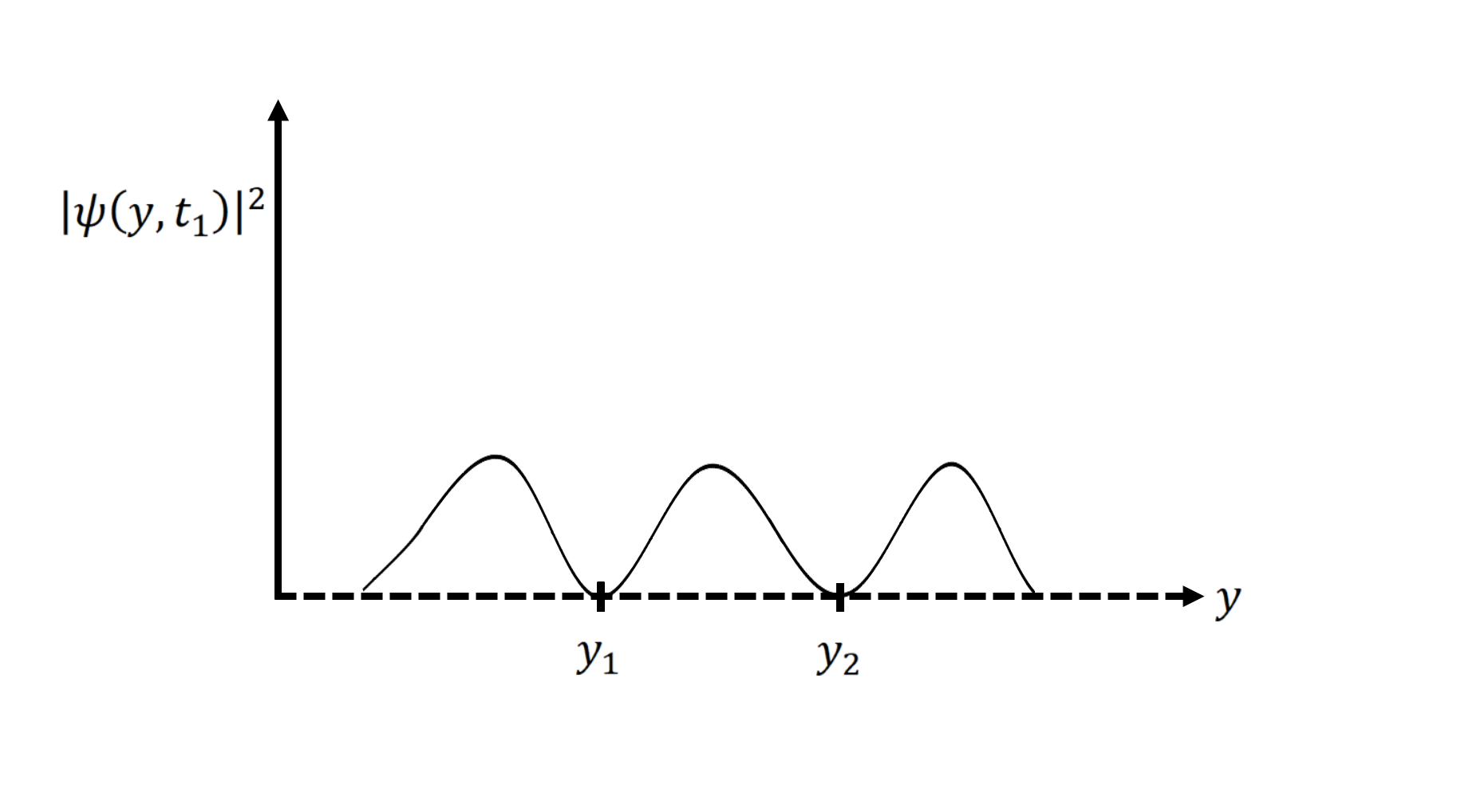}
        \caption{$t=t_1$}
        \label{fig: born violation 1 }
    \end{subfigure}
    \hfill
    \begin{subfigure}{0.49\textwidth}
        \includegraphics[width = \linewidth]{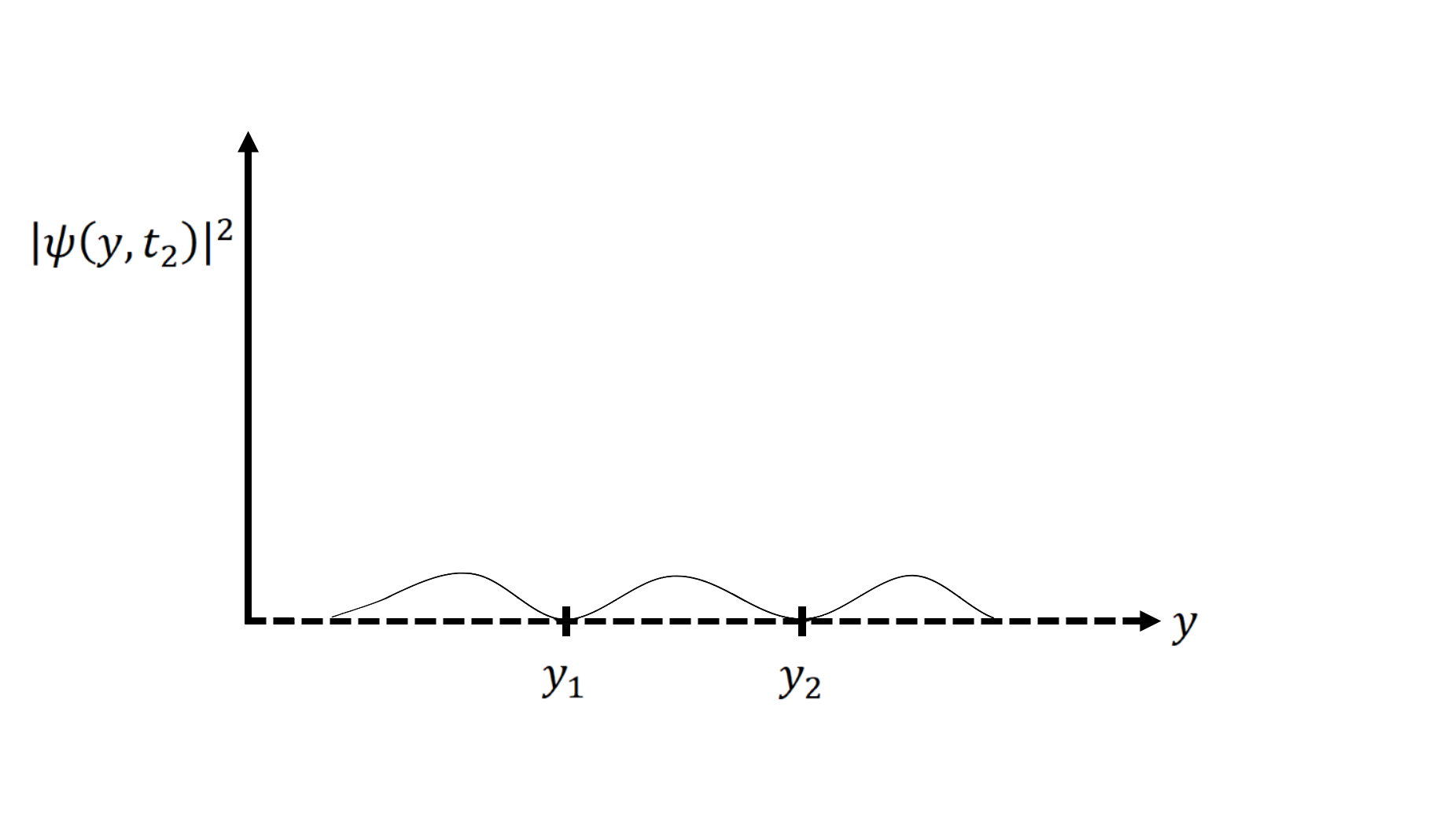}
        \caption{$t=t_2$}
        \label{fig: born violation 2 }
    \end{subfigure}
    \caption{An illustration where the $\vert\psi\vert^2$ is changing between the null, as appears to be happening in Fig. \ref{fig: bubble heat}}
    \label{fig: born violation}
\end{figure*}

We proceed by dividing the wave-function into two parts:
\begin{equation}
    \psi=\psi_l+\psi_r
\end{equation}
Here, $\psi_l$  and $\psi_r$ represents the wave-function contributions from the left and right sides, respectively with respect to the center of the view plane. For one side $x_2$ will be greater than $\alpha_i$ and for the other side, it will be smaller than $\alpha_i$, thus suggesting that respective vectors of $\psi_l$ and $\psi_r$ are in the opposite quadrants of the parametric plot. In such a case, Null can exist, as it is possible to arrange the four vectors to give a zero (as illustrated in Fig. \ref{fig: cornu mid null})


\section{Bubble and Continuity}

We were able to show that two sides are permeable, i.e., sides a and d in Fig. \ref{fig: bubble heat null}. However, the permeability of the other two sides namely b and c is still indecisive. We will carry out the analysis by assuming that the probability densities are zero along the sides b and c.

An example case has been schematically illustrated in Fig. \ref{fig: born violation}, in which the probability density at two different times $t_1$ and $t_2$ has been shown. We assume that $\left\vert\psi\left(y_1\left(t\right)\right)\right\vert^2=\left\vert\psi\left(y_2\left(t\right)\right)\right\vert^2=0$ for all the values of $t$. At time $t=t_1$,
\begin{equation}
    \int_{y_1}^{y_2}\left\vert\psi\left(y,t\right)\right\vert^2\, dy=C_1,\ t = t_1
\end{equation}
As per the Quantum equilibrium hypothesis or Born’s rule, $\rho=\left\vert\psi\right\vert^2$
at any given position and time. If the quantity $\rho$ is identified with the particle density (for example considering ensemble interpretation of the current \cite{shankar2012principles}), it must follow the continuity equation.
\begin{equation}
    \frac{\partial\rho}{\partial t}=-\mathbf{\nabla}\cdot\left(\rho \mathbf{v}\right)
\end{equation}
for which it has been shown in Sec. \ref{property trap traj} that:
\begin{equation}\label{eqn: rho unchanged}
    \frac{d}{dt}\left(\int_{V_t}\rho\, d V\right)=0
\end{equation}
Here, $V_t$ is the boundary which can be a function of time, but $\rho$ is zero on the time variable boundary.
However, at another time $t=t_2$,
\begin{equation}
    \int_{y_1}^{y_2}\left\vert\psi\left(y\right)\right\vert^2\, dy=C_2,\ t=t_2
\end{equation}
If $C_1 \ne C_2$, which is apparent from Fig. \ref{fig: bubble heat} and illustrated in the Fig. \ref{fig: born violation}, implies that:
\begin{equation}
    \frac{d}{dt}\left(\int_{y_1}^{y_2}\left\vert\psi\left(y\right)\right\vert^2\, d y\right)\neq0    
\end{equation}
which contradicts Eqn. \ref{eqn: rho unchanged}. One conclusion can be that Quantum Equilibrium Hypothesis or Born’s rule cannot hold in the given scenarios, i.e.,
\begin{equation}
    \left\vert\psi\left(y,t\right)\right\vert^2\neq\rho\left(y,t\right)
\end{equation}
However, the continuity equation is directly derivable from the Schr\"odinger equation, as discussed in section \ref{property trap traj}, which means that no situation satisfying the Schr\"odinger equation can violate the equation of continuity. Further, it is well known that Feynman Path Integral formalism and Schr\"odinger's wave mechanics are equivalent\cite{feynman2010quantum, shankar2012principles}, thus, only logical conclusion is that the side boundaries are not closed as well.

In fact, a careful numerical analysis of the above scenario reveals that the probability density at the Null boundary is quite small but finite and about six orders of magnitude less than the probability density at the peak of the bubble.

\begin{figure}
    \centering
    \includegraphics[width = 0.8\textwidth]{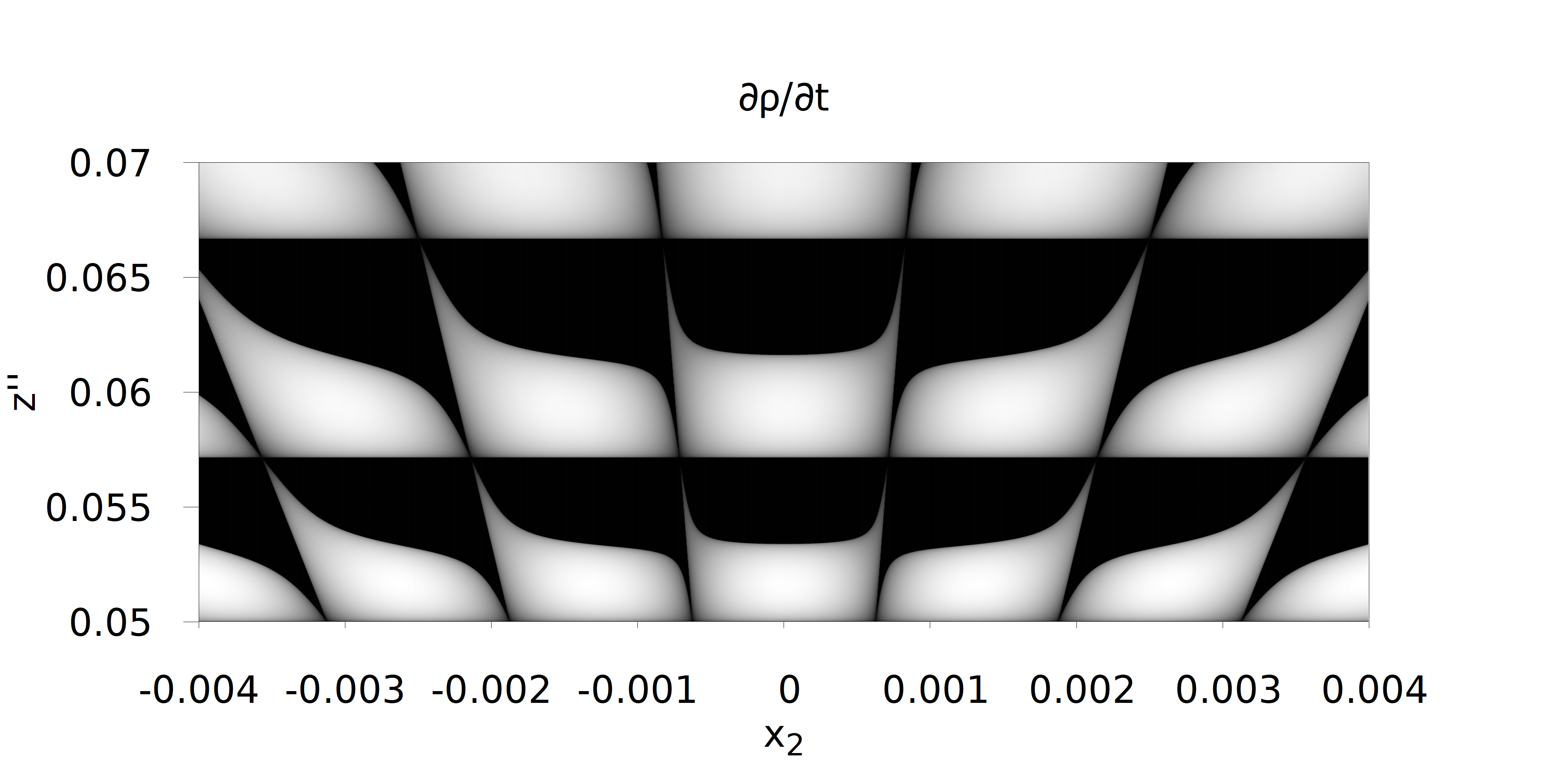}
    \caption{Heat plot of $\partial \rho / \partial t$, which is a continuous function is shown at the same location as presented in Fig \ref{fig: bubble and continuity}. The black region represents the negative values and light region is positive valued. The transition boundary of the two region represents stationary probability density points. It is apparent from this plot that the Null boundaries are stationary probability density boundaries as well.}
    \label{fig: del_rho }
\end{figure}

This prompts one to explore the flow of probability density at the null boundary. The glimpse of the change with time is shown in Fig \ref{fig: del_rho }, where, the light region represent the increment in probability density with time, whereas the dark region represents the decrement in probability density. It can be observed that the null lies at the transition of the two regions and can be stated to have stationary probability density. However, this region clearly requires more detailed analysis, for example probing the behaviour of current in the region, which is presently underway and will be reported later.

\section{Summary and Conclusions}
Although the scattering of matter waves through slits is not a new phenomenon and has been discussed and studied extensively throughout the history. However, closer inspection of probability density near the slit plane reveals intriguing structures, which seems to not to have been studied in detail in the literature. 

To study such a rich structure, we seek to look at the evolution of probability densities through a different perspective. We plot regions with near-zero probability density in the quest to get any insights into the structures found near the slits. Upon plotting Null maps, peculiar structures were revealed in the multi-slit scenario. Most notably, braids were observed in the presence of double slit, whose complexity was found to increase with increasing number of slits.

The Null map appears to have a transition zone where braided structure disappears and fringe like structure makes its appearance as seen in Fig. \ref{fig: 2 Slit SW 0.1 ISD 40 Log }. It has been demonstrated explicitly that the origin of transition zone can be understood on the basis of the hypergeometric structure of the wave-function. By making use of the oscillatory nature of the hypergeometric function, the transition region has been estimated, which is in agreement with the one obtained by explicit numerical integration.

Upon a closer examination probability density in the near slit region, we found existence of regular structures separated by boundaries of zero or near zero probability densities, which we called bubbles. As per the trajectory picture, the probability density should be trapped in a bubble, as it must follow the continuity equation. A detailed analysis using Cornu spiral revealed that the two edges (namely a and d in Fig. \ref{fig: bubble and continuity}) have to be permeable. A simple argument based on the continuity equation suggested that the side boundaries (namely b and c in Fig. \ref{fig: bubble and continuity}) have to be permeable as well.

The study of scattering through Null maps provides an intricate perspective to analyse certain details of the system, which are otherwise difficult to visualise. Specifically, this analysis has revealed the existence of intricate behaviour of the otherwise well studied multi-slit systems. 

In the future, it will be interesting to study evolution of probability current density around the "quasi-Bubbles" reported in this study, whose size can easily be controlled by changing parameters of the system suitably and can be made much larger than the wavelength used. The increasing "fuzziness" of the Null maps in the  case of multi-slit scenario, opens the possibility of studying the existence of complexity in such systems. Investigations along these lines are under progress.

\section*{Acknowledgments}

H.S. acknowledges financial support from the Department of Science and Technology (DST), Government of India through the DST-INSPIRE scheme. The authors wish to express
their heartfelt gratitude to late Prof. S. M. Chitre, without whose constant encouragement and
support this work would not have become possible. This paper is dedicated to him. We would like express a heartfelt gratitude to our anonymous referees for their critical assessment of the paper and their constructive suggestions, which lead to substantially improved quality of the study.

\bibliographystyle{apalike}






\end{document}